\documentclass[reprint,nofootinbib,amsmath,amssymb,aps,pra]{revtex4-2}

\usepackage{amsmath,amssymb,amsfonts}
\usepackage{mathtools}

\usepackage{graphicx}

\usepackage{xcolor}

\usepackage{xspace}

\usepackage{comment}

\usepackage{booktabs}
\usepackage{tabularx}
\usepackage{float}
\usepackage{bbm}

\usepackage{physics}

\usepackage[colorlinks=true,
            linkcolor=blue,
            citecolor=blue,
            urlcolor=blue]{hyperref}

\renewcommand{\tr}{\mathrm{Tr}}

\begin{document}

\title{Entanglement detection from moments of the realigned state:\\application to continuous-variable quantum systems}

\author{Serge Deside}
\author{Vojt\v{e}ch Kala}
\author{Nicolas J. Cerf}
\affiliation{Centre for Quantum Information and Communication, École polytechnique de Bruxelles, CP 165, Université libre de Bruxelles, 1050 Brussels, Belgium}

\begin{abstract}
Certifying entanglement is a central problem in quantum information sciences, especially for continuous-variable quantum systems where full state tomography is experimentally demanding. We introduce a family of separability criteria based on the singular moments and weak moments of the operator obtained by applying the realignment map to the density operator.  In practice, we find that the criterion based on the second-order moments of the realigned operator is already sufficient to witness the entanglement of a wide range of quantum optical states. The realigned moments can be readily accessed using an optical scheme consisting of a linear interferometer that couples several identical copies of the state, similarly to the multicopy technique used to measure the partial-transpose moments of a state [Phys. Rev. Research~\textbf{8}, 033064 (2026)]. We illustrate the detection capability of the second-order realigned-moment method for both Gaussian and non-Gaussian states, including lossy NOON states.
\end{abstract}

\maketitle

\section{Introduction}

Deciding whether a given quantum state is separable or entangled remains a central challenge in modern-day quantum information sciences. Since the introduction of the seminal Positive-Partial-Transpose (PPT) criterion, relying on the positivity of the partially transposed state~\cite{Peres1996,Horodecki1996}, a huge body of work has been dedicated to entanglement detection~\cite{Guehne2009,Horodecki2009}.
The problem is especially pronounced for continuous-variable (CV) quantum systems, where the underlying Hilbert space is infinite-dimensional and full state tomography becomes experimentally hard~\cite{Braunstein2005b,Lvovsky2009,Weedbrook2012}. This has motivated the development of separability criteria based on experimentally accessible quantities for CV quantum optical states, including quadrature second-order moments~\cite{Duan2000,Simon2000}, quadrature entropies~\cite{Walborn2009}, or matrices of higher-order moments of the field operators~\cite{Shchukin2005,Miranowicz2009}, to cite a few examples.

Recently, separability criteria based on the moments of the partially transposed state have been introduced~\cite{Elben2020,Yu2021,Neven2021}. These criteria, which require local random measurements followed by classical postprocessing, have been proven effective in the context of many-body systems, with a significant advantage over state tomography methods in terms of statistical errors~\cite{Elben2020}. Even more recently, these criteria have been shown to be particularly well suited for entanglement detection in CV systems, exhibiting a strong detection capability for non-Gaussian entangled states~\cite{Deside2026}. Indeed, by exploiting the multicopy technique~\cite{Brun2004}, a practical measurement scheme is achievable using multiple copies of the state of interest, together with passive linear-optical elements and particle-number measurements, circumventing the need for randomized measurements and postprocessing~\cite{Deside2026}.

In this paper, we apply a similar strategy to another separability criterion, known as the computable cross-norm criterion~\cite{Rudolph2003,Rudolph2005} or realignment criterion \cite{Realignment2003}. Such a criterion, based on the trace-norm of the so-called realigned operator, is able to detect (PPT) bound entanglement~\cite{Horodecki2009}, which provides it with a distinct advantage over all criteria derived from the partially transposed state. Along this path, separability criteria based on the moments of the realigned operator have recently been introduced~\cite{Zhang2022,Wang2024Realignment,Aggarwal2024Realignment,AggarwalAdhikari2024}, in analogy with those based on the partial-transpose moments. These works developed several refinements of the original realignment criterion based on realigned moments (or, more generally, on the singular values of the realigned operator) and have proven useful to detect certain bound entangled states of finite-dimensional systems. 
However, accessing these realignment-based quantities remains challenging and limited to realigned operators of finite rank. Hence, the corresponding methods are not suited to CV quantum states, commonly encountered in quantum optics.

Here, we focus on entanglement detection for CV systems, aiming at developing a dimension-independent approach. The original realignment criterion itself is not experimentally accessible for CV systems, which has previously motivated the development of the so-called \textit{weak} realignment criterion~\cite{Hertz2021}. The latter can be readily accessed with homodyne detection, but this advantage comes at the price of a lower entanglement detection capability. 
In the present work, we propose a family of experimentally tractable separability criteria that relate the \textit{singular} moments and \textit{weak} moments of the realigned operator, and are straightforwardly applicable to CV systems. Crucially, these criteria outperform the weak realignment criterion, while remaining experimentally friendly.

In analogy with the criteria exploiting the moments of the partially transposed state~\cite{Deside2026}, our realigned-moments criteria can be accessed using multiple copies of a CV system. The corresponding quantum optical schemes comprise passive linear optics, homodyne detection, and parity measurements. We analyze, in particular, the performance of the second-order criterion of this family for Gaussian states, and for a variety of non-Gaussian states, including mixed Schrödinger cat states and lossy NOON states.

In Sec.~\ref{sec:preli}, we review the original realignment criterion, as well as its connection to the weak realignment criterion. Then, in Sec.~\ref{sec:beyond}, we derive our realignment-moment separability criteria and demonstrate how they surpass the weak criterion, while retaining comparable experimental accessibility. In Sec.~\ref{sec:implementation}, we describe the quantum optical schemes for measuring the (singular and weak) moments of the realigned operator, and, in Sec.~\ref{sec:benchmarks}, we benchmark our new criteria on Gaussian states, as well as several non-Gaussian states. Finally, in Sec.~\ref{sec:discussion}, we discuss our results.

\section{Realignment criterion}

\label{sec:preli}

Similar to the Schmidt decomposition for pure states, bipartite mixed states can be expanded using the operator Schmidt decomposition~\cite{Hertz2021,ZhangDenker2024}. Any bipartite density operator $\hat \rho$ can be written as
\begin{equation}
\label{Schmidt}
    \hat\rho = \sum_i \lambda_i \, \hat A_i \otimes \hat B_i,
\end{equation}
where $0 \leq \lambda_i \leq 1$,  $\forall i$, and the operators $\hat A_i$ and $\hat B_i$ form orthonormal sets with respect to the Hilbert--Schmidt inner product, i.e.,
\begin{equation}
    \Tr\left[\hat A_i^\dagger \hat A_j\right]=\Tr\left[\hat B_i^\dagger \hat B_j\right]=\delta_{ij}.
\end{equation}
We refer to the subsystem associated with operators $\hat A_i$ as Alice and to the one associated with operators $\hat B_i$ as Bob. The Schmidt coefficients $\lambda_i$ are unique, whereas the associated operators $\hat A_i$ and $\hat B_i$ are not (since $\hat\rho$ is Hermitian, the operators can be chosen Hermitian too, although this choice is not necessary).

Using this notation, the computable cross-norm criterion~\cite{Rudolph2003,Rudolph2005}, states that every separable state satisfies
\begin{equation}
\label{eq:cross-norm-criterion}
    \sum_i \lambda_i \leq 1.
\end{equation}
Hence, if $\sum_i \lambda_i > 1$, the state must be entangled. This criterion is also known as the realignment criterion \cite{Realignment2003}, where  the term ``realignment" comes from the so-called realignment map $R$, defined by its action on product operators where the indices are reorganized as
\begin{equation}
\label{eq:def_realignment}
    R\left(\ket{i}\bra{j} \otimes \ket{k}\bra{l}\right):=\ket{i}\bra{k} \otimes \ket{j}\bra{l}.
\end{equation}
Then, using the Choi-Jamiołkowski vectorization~\cite{Jamiolkowski1972,Choi1975} 
\begin{equation}
\label{eq:ChoiJamiolkowski}
    \hat X=\sum_{ij} X_{ij}\ket{i}\bra{j}\quad\mapsto\quad\ket{X}=\sum_{ij} X_{ij}\ket{i}\otimes\ket{j},
\end{equation}
which associates vectors $\ket{A}$ and $\ket{B}$ to the corresponding operators $\hat A$ and $\hat B$, we find~\cite{Hertz2021}
\begin{equation}
\label{eq:realigned_schmidt}
    R(\hat\rho)=\sum_i \lambda_i\ketbra{A_i}{B_i^*}.
\end{equation}
Here, $\ket{B_i^*}$ denotes the vectorization of the complex conjugate of $\hat B_i$. Since the vectorized operators satisfy
\begin{equation}
    \langle{A_i}|{A_j}\rangle=\langle{B_i}|{B_j}\rangle=\delta_{ij},
\end{equation}
we see that Eq.~\eqref{eq:realigned_schmidt} is a singular-value decomposition of the realigned operator $R(\hat\rho)$, so that the Schmidt coefficients $\lambda_i$ are precisely the singular values of $R(\hat\rho)$.
Recalling the definition of the trace norm of an operator $\hat X$ as $\|\hat X\|_1:=\Tr \sqrt{\hat X^\dagger \hat X} $,
we obtain
\begin{equation}
    \norm{R(\hat\rho)}_1=\sum_i \lambda_i.
\end{equation}
The separability criterion \eqref{eq:cross-norm-criterion} is thus equivalent to the so-called \textit{realignment criterion}, namely
\begin{equation}
\label{eq:realignment_criterion}
    \hat\rho \ \mathrm{separable}\quad\Longrightarrow\quad\left\|R(\hat\rho)\right\|_1\leq 1.
\end{equation}

Although the trace norm of the realigned operator $R(\hat\rho)$ provides a powerful separability criterion, it is generally not directly accessible experimentally, especially for continuous-variable systems. This has motivated the introduction of a weak form of the realignment criterion, which is based on the trace (instead of the trace norm) of the realigned operator and can easily be measured~\cite{Hertz2021}.

Remember that, for every trace-class operator $\hat X$, Hölder's inequality gives~\cite{Simon2005TraceIdeals}
\begin{equation}
\left|\Tr[\hat X]\right|=\left|\Tr[\hat X\hat\openone]\right|\leq\left\|\hat X\right\|_1\left\|\hat\openone\right\|_\infty=\left\|\hat X\right\|_1.
\end{equation}
Applying this inequality to $\hat X \equiv R(\hat\rho)$ and combining it with Eq.~\eqref{eq:realignment_criterion} yields the  \textit{weak realignment criterion}, namely~\cite{Hertz2021}
\begin{equation}
\label{eq:weak_criterion}
    \hat\rho \ \mathrm{separable}\quad\Longrightarrow\quad\left|\Tr\left[R(\hat\rho)\right]\right|\leq 1.
\end{equation}
Note that whenever $\Tr[R(\hat\rho)]$ is used instead of  $\norm{R(\hat\rho)}_1$, we assume that the Hilbert spaces of Alice and Bob satisfy $\mathcal H_A \simeq \mathcal H_B \simeq \mathcal H$. Consequently, $\mathcal H_A\otimes\mathcal H_A\simeq \mathcal H_B\otimes\mathcal H_B$, so that the vectorized operators $\ket{A_i}$ and $\ket{B_i^*}$ belong to the same Hilbert--Schmidt space and $R(\hat\rho)$ can be viewed as a square operator, whose trace is thus well defined. Using Eq.~\eqref{eq:realigned_schmidt}, this trace can be written as
\begin{equation}
    \Tr\left[R(\hat\rho)\right]=\sum_i\lambda_i\langle{B_i^*}|{A_i}\rangle.
\end{equation}

Notably, the original and weak forms of the realignment criterion coincide for \textit{Schmidt-symmetric} states~\cite{Hertz2021}, i.e.,  states that admit an operator Schmidt decomposition with $\hat B_i=\hat A_i^*$, $\forall i$, hence
\begin{equation}
\label{eq:Schmidt-symm}
    \langle{B_i^*}|{A_j}\rangle=\delta_{ij}, \quad \forall i,j.
\end{equation}
Indeed, for such states, we have
\begin{equation}
    \Tr\left[R(\hat\rho)\right]=\sum_i \lambda_i=\left\|R(\hat\rho)\right\|_1.
\end{equation}
Outside this Schmidt-symmetric class, however, the weak criterion may be strictly weaker than the original realignment criterion as there exist states whose entanglement is certified by the original criterion ($ \left\|R(\hat\rho)\right\|_1 >1$) but remains undetected by the weak criterion ($\Tr\left[R(\hat\rho)\right]\le 1$). The aim of the realigned-moment criteria introduced in this work is to partially bridge this gap, while keeping the quantities needed to apply the criteria experimentally measurable.

\section{Realigned-moment criteria}

\label{sec:beyond}

We now attempt to build improved separability criteria based on the moments of the realigned operator $R(\hat{\rho})$. We consider two distinct families realigned moments. First, we consider the $n$-th \textit{singular-value moment} --  singular moment for short -- of the realigned operator $R(\hat\rho)$, namely 
\begin{equation}
    r_n := \Tr\left[\left(R(\hat{\rho})^\dagger R(\hat{\rho})\right)^{n/2}\right] = \sum_i \lambda_i^n,
\end{equation}
which corresponds to the $n$-th order moment of the singular values of $R(\hat\rho)$. These are simply called realigned moments in refs.~\cite{Zhang2022,Wang2024Realignment,Aggarwal2024Realignment,AggarwalAdhikari2024}.  Second, under the same identification of local spaces as in the previous section, we define the $n$-th \textit{weak moment} of $R(\hat{\rho})$ as
\begin{equation}
     w_n := \Tr\left[R(\hat{\rho})^n\right],
\end{equation}
since it is directly related to the weak realignment criterion.

Note that these moments satisfy $w_n\geq0$, $r_n\geq0$, and $w_n\leq r_n$, as well as $r_n\geq r_{n+1}$, for all $n\in\mathbb{N}_{>0}$ (see Appendix~\ref{app:properties of wn and rn} for the proofs). Note also that these two families of $n$-th order moments coincide ($r_n=w_n$) if $R(\hat\rho)$ is Hermitian. With this notation, the realignment and weak realignment criteria are simply rephrased, respectively, as $r_1\leq1$ and $w_1\leq1$ for all separable states.
In particular, the first-order ($n=1$) singular and weak realignment moments are given, respectively, by

\begin{equation}
  \begin{aligned}
  r_1&=\norm{R(\hat{\rho})}_1=\sum_i \lambda_i, \\ w_1&=\Tr\left[R(\hat\rho)\right]=\sum_i \lambda_i \langle{B_i^*}|{A_i}\rangle.
  \end{aligned}
\end{equation}
In order to shorten the notation, we define the cross-Gram matrix $M$ between the sets $\{\ket{A_i}\}$ and $\{\ket{B_i^*}\}$ as
\begin{equation}
    M_{ij}\coloneqq\langle B_i^*|A_j\rangle=\Tr\left[\hat B_i^T \hat A_j\right],
\end{equation}
so that we may rewrite
\begin{equation}
\label{eq:w1}
    w_1=\sum_i \lambda_i \, M_{ii}.
\end{equation}
More details on the cross-Gram matrix associated with the operator Schmidt decomposition is provided in Appendix \ref{app:A in B}. For a Schmidt-symmetric state, the matrix $M$ is just the identity matrix $I$, so its diagonal elements are equal to one, hence $w_1=r_1$.

Otherwise, the Schmidt coefficients $\lambda_i$  and their sum $r_1$ are not directly accessible from $w_1$, since the latter also depends on the overlaps
$M_{ii}=\mathrm{Tr} [\hat B_i^T \hat A_i]$. In that case, the weak realignment criterion may become inefficient compared to the original criterion. This is easily illustrated with the example of a balanced NOON state
\begin{equation}
\label{eq:noon}
    \ket{\psi_N}=\frac{1}{\sqrt{2}}\left(\ket{N,0}+\ket{0,N}\right),
\end{equation}
with $N\in \mathbb N_{>0}$ being the photon number. Its vector of operator Schmidt coefficients is
$\boldsymbol{\lambda}=\left(1/2,\,1/2,\,1/2,\,1/2\right)$,
with the corresponding operator sets
\begin{equation}
    \{\hat A_i\}_{i=1}^4=\left\{\dyad{0},\ketbra{0}{N},\ketbra{N}{0},\dyad{N}\right\},
\end{equation}
and
\begin{equation}
    \{\hat B_i\}_{i=1}^4=\left\{\dyad{N},\ketbra{N}{0},\ketbra{0}{N},\dyad{0}\right\}.
\end{equation}
While the realignment criterion certifies inseparability since $r_1=4\times 1/2 = 2$, we have $w_1=0$ since the overlaps $M_{ii}=0$ for $i=1,\dots,4$. The weak realignment criterion is thus inconclusive for the NOON state.

Clearly, the problem is that the diagonal entries of the cross-Gram matrix $M$ of the operator Schmidt decomposition of the NOON state vanish (whereas $M_{ii}=1$ for a Schmidt-symmetric state). Generalizing this example, we denote as \textit{Schmidt-antisymmetric} any state such that $M_{ii}=0$ for all $i$. For any such state, the first-order weak moment of the realigned density operator satisfies $w_1=0$ and the weak realignment criterion is unable to detect entanglement. We now show that this problem can be  circumvented by considering the second-order weak moment of the realigned operator, namely
\begin{equation}
    w_2=\Tr\left[R(\hat{\rho})^2\right].
\end{equation}
Using Eq.~\eqref{eq:realigned_schmidt}, it can be rewritten as
\begin{equation}
\label{eq:w2}
    w_2=\sum_{ij}\lambda_i\lambda_jM_{ij}M_{ji}.
\end{equation}
Clearly, in the case of Schmidt-antisymmetric states, information on the Schmidt coefficients $\lambda_i$ may become accessible from $w_2$ via the off-diagonal elements of $M$, whereas no information is contained in $w_1$ as it only depends on the diagonal elements of $M$.

In order to make use of $w_2$, we need to find a criterion involving products of Schmidt coefficients.
Let us first prove the following two inequalities, valid for all states,
\begin{equation}
\label{eq:new_ineq}
    r_2 \pm w_2\leq r_1^2 \pm w_1^2.
\end{equation}
Using Eqs.~\eqref{eq:w1} and~\eqref{eq:w2}, together with the fact that $r_1=\sum_i\lambda_i$ and $r_2=\sum_i\lambda_i^2$, we obtain
\begin{equation}
\label{eq:develop_criterion}
    r_1^2 \pm w_1^2 - (r_2 \pm w_2)
    =\sum_{i\neq j}\lambda_i\lambda_j\left[1\pm\left(M_{ii}M_{jj}-M_{ij}M_{ji}\right)\right].
\end{equation}
In Appendix~\ref{app:w2bound}, we prove that
\begin{equation}
    \left|M_{ii}M_{jj}-M_{ij}M_{ji}\right|\leq1.
\end{equation}
The quantities $d_{ij}:=M_{ii}M_{jj}-M_{ij}M_{ji}$ need not be real, but since the left-hand side of Eq.~\eqref{eq:develop_criterion} is real (see Appendix~\ref{app:properties of wn and rn}), we may take the real part of $d_{ij}$ and obtain
\begin{equation}
    r_1^2\pm w_1^2-(r_2\pm w_2)=\sum_{i\neq j}\lambda_i\lambda_j\left[1\pm\mathrm{Re}(d_{ij})\right].
\end{equation}
The bound $|d_{ij}|\leq 1$ implies $|\mathrm{Re}(d_{ij})|\leq 1$, so every term in the last sum is non-negative. This proves Eq.~\eqref{eq:new_ineq}.

From the realignment criterion, separable states satisfy $r_1\leq1$, and thus also $r_1^2\leq1$. Therefore, Eq.~\eqref{eq:new_ineq} implies the following \textit{second-moment realignment criterion}
\begin{equation}
\label{eq:our_criterion}
    \hat{\rho}\ \mathrm{ separable}\implies r_2 + |w_2-w_1^2|\leq1.
\end{equation}
Equivalently, entanglement is certified whenever
\begin{equation}
\label{eq:criterionw2}
    r_2 + |w_2-w_1^2|>1.
\end{equation}
Moreover, since $w_2\leq r_2$, any violation of the weak realignment criterion ($w_1>1$) also implies
\begin{equation}
    r_2 + (w_1^2-w_2)>1.
\end{equation}
In this sense, the second-moment realignment criterion~\eqref{eq:our_criterion} contains (and is stronger than) the weak realignment criterion~\eqref{eq:weak_criterion}. 
Let us now reexamine the NOON state and check that criterion~\eqref{eq:our_criterion} works. Since all Schmidt coefficients are 1/2, we have $r_2=4\times 1/4=1$, and since the state is Schmidt-antisymmetric, we have $w_1=0$. Further, $r_2=w_2$ because $R(\hat{\rho})$ is Hermitian in this case. Hence,
$r_2+|w_2-w_1^2|=2$, and entanglement is certified. In Sec.~\ref{sec:benchmarks}, we will pursue this analysis for lossy NOON states as well as other typical quantum optical states in order to benchmark the method. 

Interestingly, the second-order singular moment $r_2=\sum_i \lambda_i^2$ happens to be equal to the purity $\tr\left[\hat \rho^2\right] $ of the state $\hat\rho$, since, by using the Hermiticity of $\hat\rho$, we get
\begin{equation}
\label{eq:purity=r2}
    \tr\left[\hat \rho ^\dagger \hat \rho\right] =\sum_{ij} \lambda_i \lambda_j \Tr[A_i^\dagger A_j] \Tr[B_i^\dagger B_j] = \sum_i \lambda_i^2 .
\end{equation}
As a consequence, we remark that the second-moment realignment criterion~\eqref{eq:our_criterion} detects all pure entangled states. Indeed, for a pure state (i.e., $r_2=1$), entanglement is detected as soon as $|w_2-w_1^2|>0$, or, equivalently, $\sum_{i\ne j} \lambda_i \lambda _j \,|d_{ij}|>0$. The latter quantity vanishes only if the Schmidt rank is one, that is, if all Schmidt coefficients vanish except a single one, in which case we have a product state $\hat \rho = \hat A \otimes \hat B$.

The first- and second-order weak moments $w_1$ and $w_2$ also enjoy a simple form, which can be obtained by exploiting the expression of realigned operator in terms of the partial transposition $T_B$ on Bob's side, namely~\cite{Hertz2021}
\begin{equation}
\label{eq:alternative-def-R-with-T_B}
R(\hat{\rho})=(\hat{\rho}^{T_B}\hat{S}_{AB})^{T_B},
\end{equation}
where $\hat{S}_{AB}=\sum_{i,j}\ket{i}\bra{j}\otimes\ket{j}\bra{i}$ denotes the swap operator (between Alice and Bob). Using the identities $\tr(\hat X^{T_B})=\tr(\hat X)$ and $\tr(\hat X^{T_B}\hat Y^{T_B})=\tr(\hat X\hat Y)$ for any bipartite operators $\hat X$ and $\hat Y$, we have
\begin{equation}
w_1 = \tr[R(\hat\rho)] = \tr[\hat{\rho}^{T_B} \hat{S}_{AB}] = \tr[\hat{\rho} \,(\hat{S}_{AB})^{T_B}] .
\end{equation}
Noticing that $(\hat{S}_{AB})^{T_B} = \ket{\Omega}\bra{\Omega}$, where 
\begin{equation}
\label{eq:epr-proj}
    |\Omega\rangle=\sum_i |i\rangle\otimes |i\rangle.
\end{equation}
denotes the (unnormalized) maximally-entangled EPR-like state, 
we obtain~\cite{Hertz2021}
\begin{equation}
\label{eq:nice-expression-w1}
  w_1 = \tr[\hat{\rho} \, \ket{\Omega}\bra{\Omega}] = \bra{\Omega} \hat{\rho} \ket{\Omega}.
\end{equation}
In other words, $w_1$ can ve viewed as the overlap of state $\hat\rho$ with respect to the (unnormalized) EPR-like state $|\Omega\rangle$. Similarly, using Eq.~\eqref{eq:alternative-def-R-with-T_B}, we have
\begin{equation}
w_2 = \tr[R(\hat\rho)^2] = \tr[\hat{\rho}\,(\hat{S}_{AB}\hat{\rho}^{T_B}\hat{S}_{AB})^{T_B}]  .
\end{equation}
It is easy to check that, for any bipartite operator $\hat X$,
\begin{equation}
(\hat{S}_{AB}\, \hat{X}^{T_B}\hat{S}_{AB})^{T_B} = \hat{S}_{AB}\, \hat{X}^{T} \, \hat{S}_{AB},
\end{equation}
so that we obtain
\begin{equation}
\label{eq:nice-expression-w2}
w_2 =  \tr[\hat{\rho}\, \hat{S}_{AB}\, \hat\rho^T \, \hat{S}_{AB}]  .
\end{equation}
Thus, $w_2$ can be viewed as the overlap between state $\hat\rho$ and the operator obtained by swapping the roles of Alice and Bob on the transpose of $\hat\rho$. Another expression for $w_2$ can be written in terms of EPR-like states $|\Omega\rangle$, but we defer it until the discussion of the optical realization (Sec.~\ref{sec:implementation}), where it arises more naturally.

In contrast with the original realignment criterion, 
the second-moment realignment criterion~\eqref{eq:our_criterion} is experimentally accessible. As we will see in Sec.~\ref{sec:implementation}, the first and second weak moments $w_1$ and $w_2$ can readily be accessed using the multicopy method, which boils down to using a linear interferometer supplemented with homodyne measurements. Similarly, the purity $r_2$ can also be measured using the multicopy technique supplemented with photon-number measurement \cite{Griffet2023a}. Furthermore, even in case $r_2$ is not accessible, it remains possible to write a weaker criterion that only depends on $w_1$ and $w_2$. Indeed, recalling that $w_2\le r_2$, we obtain the \textit{relaxed second-moment realignment criterion}
\begin{equation}
    \hat{\rho}\ \mathrm{ separable}\implies w_2 + |w_2-w_1^2|\leq1.
\end{equation}

The above development for the second moments can be extended to a whole family of higher-order criteria (see Appendix~\ref{app:more than 2}).  Whereas all weak moments $w_n$ can, in principle, be measured using the multicopy method, only even-order singular moments $r_{2n}$ can be measured similarly, so that additional bounds are required to obtain experimentally accessible criteria at orders higher than two (see Appendix~\ref{app:general-multicopy}).

\begin{figure}[h!]
\centering
\includegraphics[width=1.0\columnwidth]{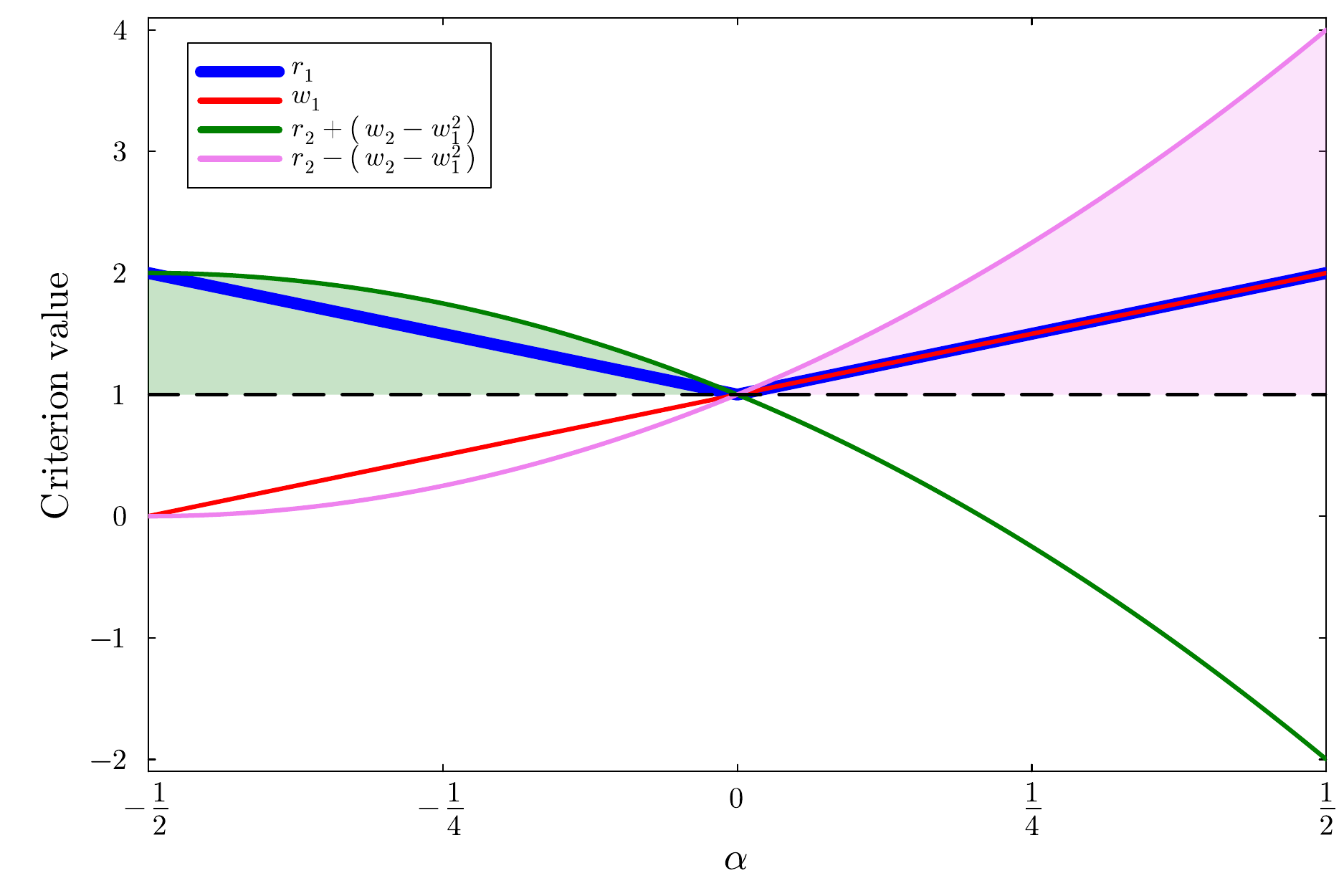}
\caption{Separability criteria for the state $\ket{\psi_\alpha}$ from Eq.~\eqref{eq:superposition}. As soon as a value greater than 1 is reached, entanglement is certified. We observe that the original realignment criterion (blue curve), based solely on $r_1$, flags entanglement for all $\alpha\neq0$, whereas the weak realignment criterion (red curve), based solely on $w_1$, flags entanglement only in the region $\alpha>0$ (notice that in this region, the blue and red curves coincide). The second-moment realignment criterion (green and pink curves), based on $r_2$, $w_2$, and $w_1$, detects entanglement for the whole entangled parameter range $\alpha\ne 0$, so it performs as well as the original realignment criterion.}
\label{fig:superposition}
\end{figure}

\begin{figure}[ht!]
\centering
\includegraphics[width=1.0\columnwidth]{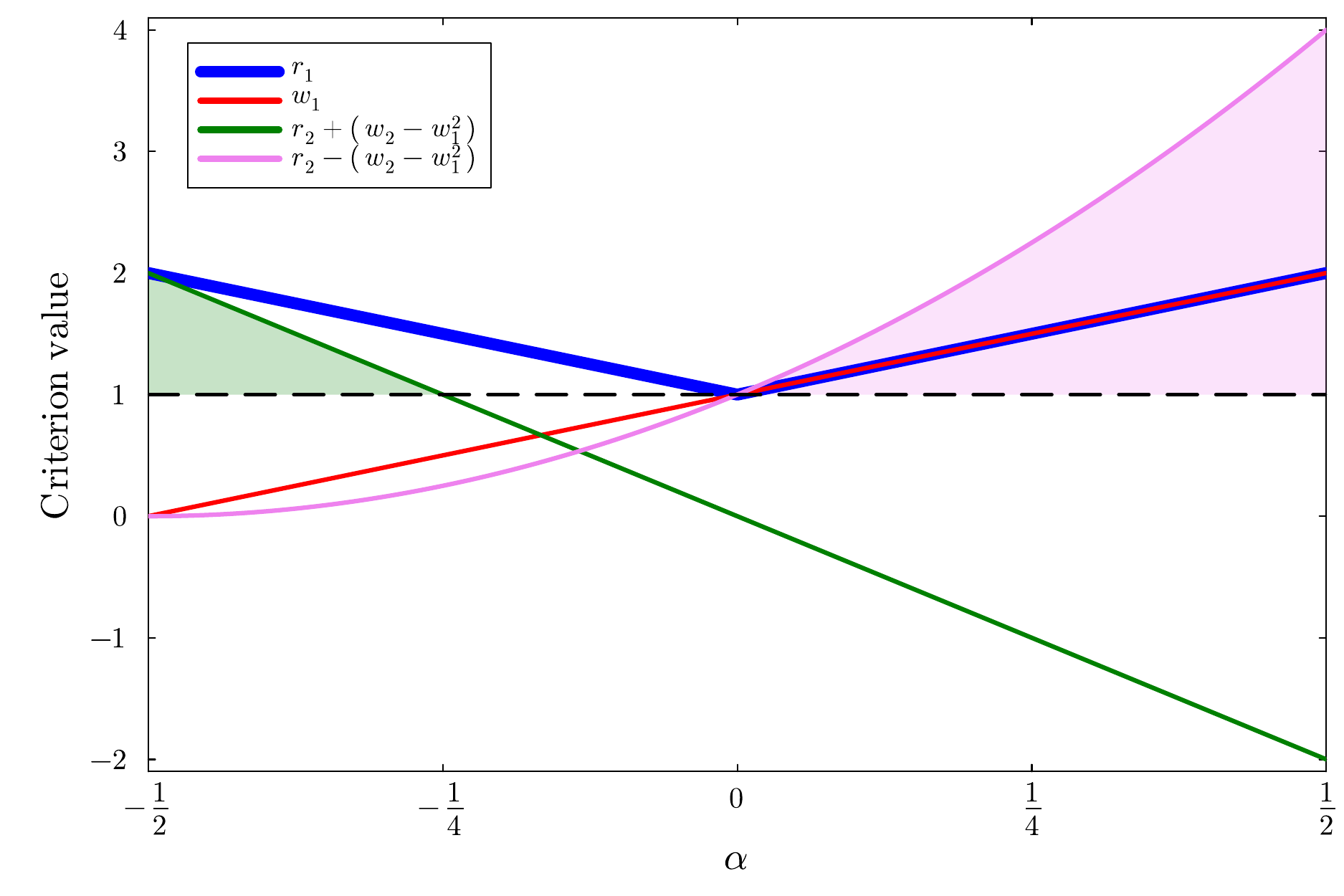}
\caption{Separability criteria for the state $\hat\rho_\alpha$ from Eq.~\eqref{eq:mixture}. As soon as a value greater than 1 is reached, entanglement is certified. We observe that the original realignment criterion (blue curve), based solely on $r_1$, flags entanglement for all $\alpha\neq0$, whereas the weak realignment criterion (red curve), based solely on $w_1$, flags entanglement only in the region $\alpha>0$ (notice that in this region, the blue and red curves coincide). The second-moment realignment criterion (green and pink curves), based on $r_2$, $w_2$, and $w_1$, detects entanglement for $\alpha \in [-1/2,-1/4) \cup (0,1/2]$. Its performance is thus halfway between the original and the weak realignment criteria.}
\label{fig:mixture}
\end{figure}

We conclude this section by analyzing the capability of the second-moment criterion~\eqref{eq:our_criterion} for states interpolating between a Schmidt-antisymmetric and Schmidt-symmetric state. In addition to the Schmidt-antisymmetric state from Eq.~\eqref{eq:noon} with $N=1$, which we denote by $\ket{\psi_a}$, we consider the Schmidt-symmetric state
\begin{equation}
\label{eq:symmetric}
    \ket{\psi_s}=\frac{1}{\sqrt{2}}\left(\ket{0,0}+\ket{1,1}\right).
\end{equation}
The operator Schmidt decomposition of its density operator $\dyad{\psi_s}$ gives
$\boldsymbol{\lambda}=\left(1/2,\,1/2,\,1/2,\,1/2\right)$,
with
\begin{equation}
    \{\hat A_i\}_{i=1}^4=\left\{\dyad{0},\ketbra{0}{1},\ketbra{1}{0},\dyad{1}\right\},
\end{equation}
and
\begin{equation}
    \{\hat B_i\}_{i=1}^4=\left\{\dyad{0},\ketbra{0}{1},\ketbra{1}{0},\dyad{1}\right\}.
\end{equation}
As expected, $M=I$. We then consider the superposition
\begin{equation}
\label{eq:superposition}
    \ket{\psi_\alpha}=\sqrt{\frac{1}{2}-\alpha}\ket{\psi_a}+\sqrt{\frac{1}{2}+\alpha}\ket{\psi_s},
\end{equation}
where $\alpha$ ranges from $-\frac{1}{2}$, corresponding to the Schmidt-antisymmetric case, to $\frac{1}{2}$, corresponding to the Schmidt-symmetric case. We have 
\begin{equation}
    r_1 = 1+ 2|\alpha|, \quad w_1 = 1+2\alpha, \quad w_2 = r_2 = 1.
\end{equation}
and the dependence of the criteria on $\alpha$ is displayed in Fig.~\ref{fig:superposition}. Entanglement is thus detected in the entire entangled parameter range $\alpha\ne 0$, as expected since $\ket{\psi_\alpha}$ is a pure state.
Similarly, we evaluate the criteria for a mixture of the Schmidt-antisymmetric and Schmidt-symmetric states
\begin{equation}
\label{eq:mixture}
    \hat\rho_\alpha=\left(\frac{1}{2}-\alpha\right)\dyad{\psi_a}+\left(\frac{1}{2}+\alpha\right)\dyad{\psi_s}.
\end{equation}
Here again, $\alpha\in[-1/2,1/2]$, and we have 
\begin{equation}
    r_1 = 1+ 2|\alpha|, \quad w_1 = 1+2\alpha, \quad w_2 = r_2 = \frac{1}{2} + 2\alpha^2,
\end{equation}
and the results are shown in Fig.~\ref{fig:mixture}. Entanglement is detected in a parameter range $\alpha \in[-1/2, -1/4]$ where the weak realignment criterion fails, but not in the entire entangled parameter range $\alpha\ne 0$ as detected by the original realignment criterion.

\begin{figure}[ht!]
\centering
\includegraphics[width=1.0\columnwidth]{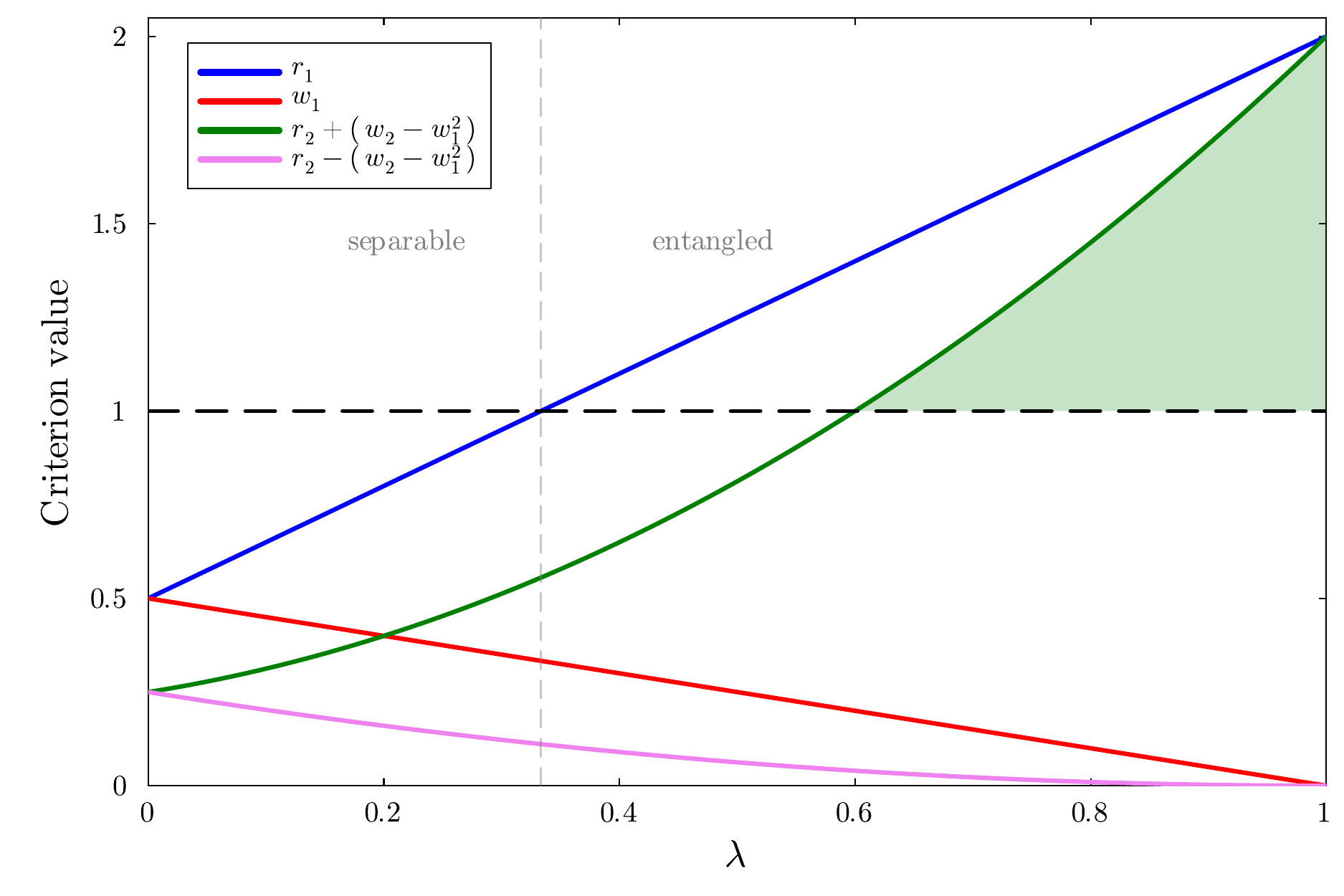}
\caption{The Werner state is entangled as soon as $\lambda>1/3$, as detected by the original realignment criterion (based on $r_1$, blue curve). The weak realignment criterion (based on $w_1$, red curve) never detects entanglement, whereas the second-moment realignment criterion (based on $r_2$, $w_2$, and $w_1$, green and pink curves) detects entanglement in the region $\lambda > 0.6$, i.e., when the state is sufficiently pure.}
\label{fig:werner}
\end{figure}

Another insightful illustration of the second-moment realignment criterion is provided by the Werner state of two qubits in the Fock basis, namely
\begin{align}
    \hat{\rho}_W = \lambda \ket{\psi^-}\bra{\psi^-}+\frac{1-\lambda}{4}\, \hat{\mathbbm{1}}, \quad \lambda \in [0,1],
\end{align}
where $\ket{\psi^-} =  (\ket{0,1} - \ket{1,0})/\sqrt{2}$. 
Figure~\ref{fig:werner} displays the detected regions using the different criteria. Again, the performance of the second-moment realignment criterion is below the original realignment criterion, although, unlike the weak realignment criterion, it successfully detects the entanglement for $\lambda \geq 0.6$.

\section{Quantum optical implementation}
\label{sec:implementation}

In this section, we describe measurement schemes for the moments $r_2$, $w_1$, and $w_2$ using multicopy~\cite{Brun2004} and interferometric techniques (see Appendix~\ref{app:general-multicopy} for the multicopy operators associated with $w_n$ and $r_{2n}$, with $n \in \mathbb N_{>0}$, as well as the corresponding quantum-optical measurement schemes). For the optical implementation, we assume that subsystems $A$ and $B$ are encoded in bosonic modes with annihilation operators $\hat a_A$ and $\hat a_B$. We also assume that the realignment map $R$ acts in the Fock basis (remember that the map is basis-dependent, just like transposition or complex conjugation).

We first consider the purity of the bipartite state $\hat\rho_{AB}$, namely $r_2=\Tr[\hat\rho_{AB}^2]$.
Let $\hat\rho_{A_1B_1}\otimes \hat\rho_{A_2B_2}$ denote two identical copies of the state. The swap operator between two copies of a system $\alpha$ ($\alpha=A,B$) is denoted by $\hat S_{\alpha_1\alpha_2}$ and is defined as
\begin{equation}
    \hat S_{\alpha_1\alpha_2} \left(|i\rangle_{\alpha_1}\otimes |j\rangle_{\alpha_2}\right)=|j\rangle_{\alpha_1}\otimes |i\rangle_{\alpha_2}.
\end{equation}
Using the identity
\begin{equation}
\label{eq:swapdef}
    \Tr\left[\hat S_{\alpha_1\alpha_2}(\hat X\otimes \hat Y)\right]=\Tr[\hat X\hat Y],
\end{equation}
which is valid for all operators $\hat X$, $\hat Y$, one obtains
\begin{equation}
    r_2=\Tr\left[(\hat S_{A_1A_2}\otimes \hat S_{B_1B_2})(\hat\rho_{A_1B_1}\otimes \hat\rho_{A_2B_2})\right].
\end{equation}
Thus, $r_2$ can be accessed by measuring the local swap operators (viewed as observables) between the two copies on Alice's side and between the two copies on Bob's side.

\begin{figure*}[t]
\centering
\includegraphics[width=\textwidth]{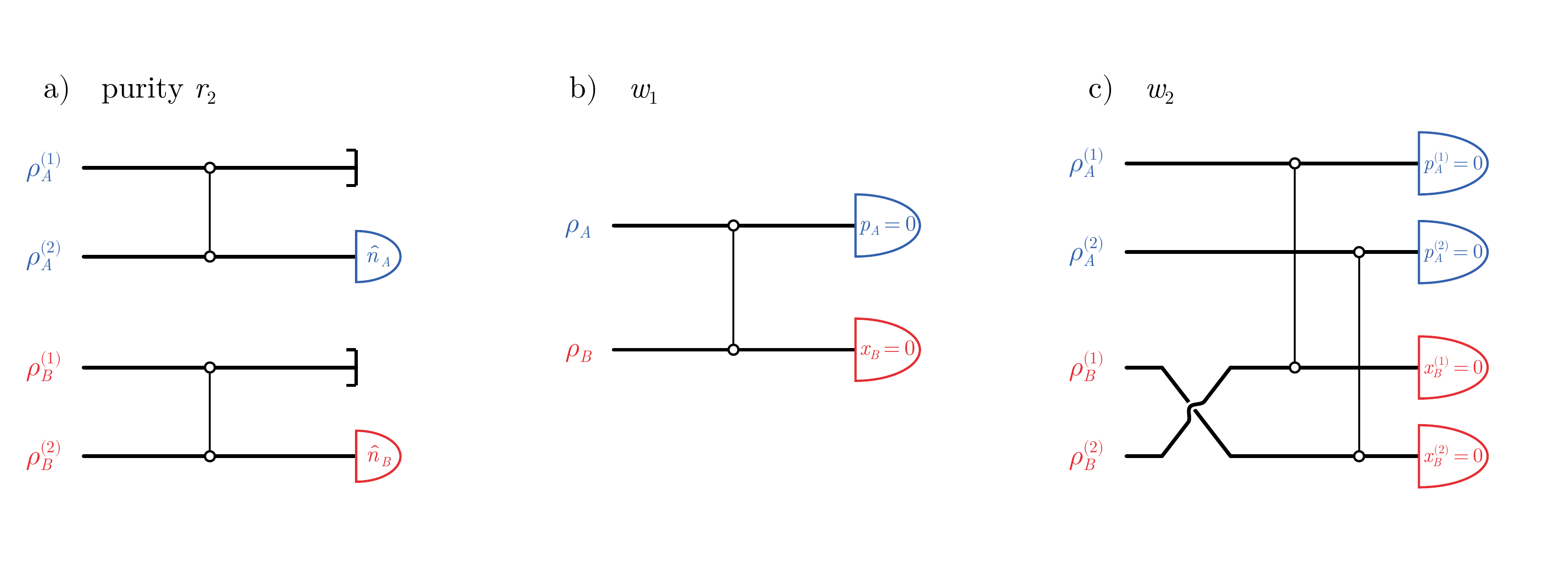}
\caption{Quantum-optical schemes for measuring the three moments entering our entanglement witness in Eq.~\eqref{eq:criterionw2}. Vertical bars represent balanced beam splitters. The first (or upper) output mode of each beam splitter corresponds to the plus sign in Eq.~\eqref{eq:beam_splitter_bis}, while the second (or lower) output mode corresponds to the minus sign in Eq.~\eqref{eq:beam_splitter_bis}. Half-ellipses denote photon-number-parity detectors in panel~a) and homodyne detectors in panels~b) and~c). \textbf{a)} The purity $r_2$ is measured by locally interfering two copies (called 1 and 2) on balanced beam splitters and measuring the photon-number parity of Alice's and Bob's second output mode. The product of the two parity outcomes is averaged over the experimental runs. \textbf{b)} The moment $w_1$ is accessed using a balanced beam splitter followed by homodyne detection of two conjugate quadratures. \textbf{c)} The moment $w_2$ is accessed by extending the scheme of panel~b) to two copies, supplemented with an exchange of Bob's two modes.}
\label{fig:implementation}
\end{figure*}

For bosonic modes, such a swap measurement can be implemented by interfering the two copies on balanced beam splitters, followed by parity detection~\cite{Islam2015,Griffet2023a}. We define the parity operator on mode $X$ as 
\begin{equation}
    \hat \Pi_{\alpha_i}=(-1)^{\hat n _{\alpha_i}},
\end{equation}
where $\hat n_{\alpha_i}=\hat a_{\alpha_i}^\dagger \hat a_{\alpha_i}$ and $i=1,2$. The balanced beam splitter between modes $\alpha_1$ and $\alpha_2$ is defined as
\begin{equation}
\label{eq:beam_splitter}
    \hat B_{\alpha_1\alpha_2}=e^{\frac{\pi}{4}\left(\hat a_{\alpha_1}^\dagger \hat a_{\alpha_2}-\hat a_{\alpha_1} \hat a_{\alpha_2}^\dagger\right)},
\end{equation}
and satisfies
\begin{equation}
\label{eq:beam_splitter_bis}
\begin{split}
&\hat B_{\alpha_1\alpha_2}^\dagger \hat a_{\alpha_1} \hat B_{\alpha_1\alpha_2}=\frac{\hat a_{\alpha_1}+\hat a_{\alpha_2}}{\sqrt2},\\&\hat B_{\alpha_1\alpha_2}^\dagger \hat a_{\alpha_2} \hat B_{\alpha_1\alpha_2}=\frac{-\hat a_{\alpha_1}+\hat a_{\alpha_2}}{\sqrt2}.
\end{split}
\end{equation}
With this convention, the swap operator is mapped onto the parity operator of the second output mode, namely,
\begin{equation}
\label{eq:swap_diagonalized}
    \hat S_{\alpha_1\alpha_2} = \hat B_{\alpha_1\alpha_2}^\dagger  (\hat{\mathbbm{1}}_{\alpha_1} \otimes \hat \Pi_{\alpha_2}) \hat B_{\alpha_1\alpha_2}.
\end{equation}
Therefore, the purity can be written as
\begin{equation}
\begin{aligned}
r_2 = \Tr\Big[
&\left(\hat{\mathbbm{1}}_{A_1} \otimes \hat \Pi_{A_2}\otimes \hat{\mathbbm{1}}_{B_1} \otimes \hat \Pi_{B_2}\right)
 \left(\hat B_{A_1A_2}\otimes \hat B_{B_1B_2}\right)
\\ \times
& \left(\hat\rho_{A_1B_1}\otimes \hat\rho_{A_2B_2}\right)
 \left(\hat B_{A_1A_2}^\dagger\otimes
       \hat B_{B_1B_2}^\dagger\right)
\Big].
\end{aligned}
\end{equation}
As a result, $r_2$ is obtained from the expectation value of the product of the photon-number parities measured in the second output modes $A_2$ and $B_2$ after locally interfering the two copies on balanced beam splitters. This is illustrated in Fig.~\ref{fig:implementation}~a).

For the measurement of $w_1$, we now exploit Eq.~\eqref{eq:nice-expression-w1}, namely 
\begin{equation}
\label{eq:reminder-w1}
\begin{aligned}
w_1=\tr[R(\hat\rho_{AB})] = \tr[\hat{\rho}_{AB} \, \ket{\Omega}\bra{\Omega}_{AB}],
\end{aligned}
\end{equation}
where we define the maximally-entangled vector
\begin{equation}
\label{eq:epr-proj}
    |\Omega\rangle_{AB}=\sum_i |i\rangle_A\otimes |i\rangle_B.
\end{equation}
In the continuous-variable optical setting, $|\Omega\rangle_{AB}$ is understood as the (unnormalized) two-mode squeezed vacuum state with infinite
squeezing. With the quadrature convention
\begin{equation}
    \hat x=\frac{\hat a+\hat a^\dagger}{\sqrt2},\qquad\hat p=\frac{\hat a-\hat a^\dagger}{i\sqrt2},\qquad[\hat x,\hat p]=i,
\end{equation}
and the corresponding quadrature eigenstates $\ket{x}$ and $\ket{p}$ normalized as
    $\langle x|x'\rangle=\delta(x-x')$ and $\langle p|p'\rangle=\delta(p-p')$,
one has~\cite{Hertz2021}
\begin{equation}
    |\Omega\rangle_{AB}=\sqrt{\pi}\,\hat B_{AB}^\dagger\left(|p{=}0\rangle_A\otimes|x{=}0\rangle_B\right).
\end{equation}
Using this expression, Eq.~\eqref{eq:reminder-w1} becomes
\begin{equation}
    w_1=\pi\Tr\!\left[\left(|p{=}0\rangle\langle p{=}0|_A\otimes|x{=}0\rangle\langle x{=}0|_B\right)\hat B_{AB}\hat\rho\hat B_{AB}^\dagger\right].
\end{equation}
Thus, after applying a balanced beam splitter $\hat B_{AB}$ to the two modes $A$ and $B$ of state $\hat\rho$, one performs homodyne detection of the $p$ quadrature in the first output mode and $x$ quadrature in the second output mode. Then, the measurement works as illustrated in Fig.~\ref{fig:implementation}~b), that is,
\begin{equation}
    w_1=\pi\,P_{p_A,x_B}(0,0),
\end{equation}
where $P_{p_A,x_B}(0,0)$ denotes the joint probability density at the origin.
The projectors onto $|x{=}0\rangle$ and $|p{=}0\rangle$ are understood in the usual distributional sense. Of course, experimentally, the joint probability density must be estimated from finite-resolution homodyne data.

The measurement of $w_2$ can be implemented similarly, using two copies of the bipartite state. We start from Eq.~\eqref{eq:nice-expression-w2} and write
\begin{equation}
\begin{aligned}
w_2& =  \tr[\hat{\rho}_{AB}\, ({\hat S}_{AB}\hat\rho_{AB}^T {\hat S}_{AB})]  \\
&=\tr[(\hat{\rho}_{A_1 B_1}\otimes {\hat S}_{A_2 B_2}\hat\rho_{A_2 B_2}^T {\hat S}_{A_2 B_2}) ({\hat S}_{A_1 A_2} \otimes {\hat S}_{B_1 B_2} ) ] \\
&=\tr[(\hat{\rho}_{A_1 B_1}\otimes \hat\rho_{A_2 B_2}^T ) ({\hat S}_{A_1 B_2} \otimes {\hat S}_{B_1 A_2} ) ] \\
&=\tr[(\hat{\rho}_{A_1 B_1}\otimes \hat\rho_{A_2 B_2} ) (({\hat S}_{A_1 B_2})^{T_{B_2}} \otimes ({\hat S}_{B_1 A_2})^{T_{A_2}} ) ] \\
&=\tr[(\hat{\rho}_{A_1 B_1}\otimes \hat\rho_{A_2 B_2} ) 
(\ket{\Omega}\bra{\Omega}_{A_1 B_2} \otimes \ket{\Omega}\bra{\Omega}_{B_1 A_2} ) ].
\end{aligned}
\end{equation}
Thus, we have 
\begin{equation}
    w_2=\Tr\left[(\hat\rho_{A_1B_1}\otimes \hat\rho_{A_2B_2})\, \hat O_2\right],
\end{equation}
where $\hat O_2$ is a two-copy observable, which can be expressed using two copies of the EPR-like state $|\Omega\rangle$ and a swap between Bob's two modes:
\begin{equation}
    \hat O_2=\hat S_{B_1B_2}\left(|\Omega\rangle\langle\Omega|_{A_1B_1}\otimes|\Omega\rangle\langle\Omega|_{A_2B_2}\right)\hat S_{B_1B_2}.
\end{equation}
Hence, using the cyclicity of the trace,
\begin{equation}
\begin{aligned}
w_2 = \Tr\Big[
&\hat S_{B_1B_2}
 \left(\hat\rho_{A_1B_1}\otimes \hat\rho_{A_2B_2}\right)
 \hat S_{B_1B_2}
\\
&\times
 \left(
 |\Omega\rangle\langle\Omega|_{A_1B_1}
 \otimes
 |\Omega\rangle\langle\Omega|_{A_2B_2}
 \right)
\Big].
\end{aligned}
\end{equation}
For compactness, we define
\begin{equation}
    \hat Q_{A_jB_j}=|p=0\rangle\langle p=0|_{A_j}\otimes|x=0\rangle\langle x=0|_{B_j},
\end{equation}
where $j=1,2$, which gives
\begin{equation}
|\Omega\rangle\langle\Omega|_{A_j B_j} = \pi \hat B^\dagger_{A_j B_j} \hat Q_{A_j B_j} B_{A_j B_j} .
\end{equation}
Substituting into the above expression gives
\begin{equation}
\begin{aligned}
&w_2 = \pi^2\,\Tr\Big[
\left(\hat Q_{A_1B_1}\otimes\hat Q_{A_2B_2}\right)
 \left(\hat B_{A_1B_1}\otimes\hat B_{A_2B_2}\right)
\\
&\times \left( \hat S_{B_1B_2}
 \left(\hat\rho_{A_1B_1}\otimes\hat\rho_{A_2B_2}\right)
 \hat S_{B_1B_2} \right)
 \left(
 \hat B_{A_1B_1}^{\dagger}
 \otimes
 \hat B_{A_2B_2}^{\dagger}
 \right)
\Big].
\end{aligned}
\end{equation}
Thus, $w_2$ can be obtained by swapping Bob's two copies, applying balanced beam splitters to the pairs $A_1B_1$ and $A_2B_2$, and measuring the joint homodyne probability density at
   $p_{A_1}=x_{B_1}=p_{A_2}=x_{B_2}=0$.
Equivalently, one may avoid a physical swap by routing the modes so that the two beam splitters mix the pairs $A_1B_2$ and $A_2B_1$ instead of $A_1B_1$ and $A_2B_2$.
The quantum-optical scheme for measuring $w_2$ is illustrated in Fig.~\ref{fig:implementation}~c).

A useful consequence of the homodyne protocol is that, for a certain class of states, it also provides an indirect determination of the purity. Remembering 
\begin{equation}
    r_2=\Tr\!\left[R(\hat\rho)^\dagger R(\hat\rho)\right]=\Tr\left[\hat\rho^2\right],
\end{equation}
we see that if the realigned operator is Hermitian in the fixed local bases used to define the realignment map, i.e., $R(\hat\rho)=R(\hat\rho)^\dagger$, then
    $r_2=\Tr\!\left[R(\hat\rho)^2\right]=w_2$.
In this case, one may estimate $w_2$ with the two-copy homodyne protocol described above and then infer the purity  without having to perform the photon-number-parity measurement that would directly access $r_2$. This yields an indirect determination of the purity via homodyning, which is valid for states such that one can establish the Hermiticity of $R(\hat\rho)$.

\section{Benchmarking the criterion}
\label{sec:benchmarks}

In this section, we investigate whether the second-moment realignment criterion~\eqref{eq:our_criterion} can be used to certify entanglement for several typical quantum-optical states. We first consider Gaussian states as a test case, and then look at important families of non-Gaussian states, namely the two-mode entangled cats, and the lossy NOON states.

\subsection{Gaussian states}

We consider an $n \times n$ continuous-variable system, where Alice and Bob each hold $n$ bosonic modes. We collect the quadrature operators in the vector
\begin{equation}
    \hat{\boldsymbol r} \coloneqq \left( \hat x_{A_1},\hat p_{A_1},\ldots,\hat x_{A_n},\hat p_{A_n}, \hat x_{B_1},\hat p_{B_1},\ldots,\hat x_{B_n},\hat p_{B_n} \right)^T.
\end{equation}
The covariance matrix $\gamma$ of a state $\hat\rho$ is defined by
\begin{equation}
    \gamma_{ij}\coloneqq\frac{1}{2}\Tr\left[\hat\rho\,\left\{\Delta \hat r_i,\Delta \hat r_j\right\}\right],\qquad\Delta \hat r_i\coloneqq\hat r_i-\Tr[\hat\rho \hat r_i],
    \end{equation}
where $\{\cdot, \cdot\}$ denotes the anti-commutator.
The displacement vector is
\begin{equation}
    \boldsymbol d\coloneqq\Tr[\hat\rho\,\hat{\boldsymbol r}].
\end{equation}
A Gaussian state is fully characterized by its first and second moments, $\boldsymbol d$ and $\gamma$. Its Wigner function is
\begin{equation}
    W_{\hat\rho}(\boldsymbol r)=\frac{1}{(2\pi)^{2n}\sqrt{\det\gamma}}e^{-\frac{1}{2}(\boldsymbol r-\boldsymbol d)^T\gamma^{-1}(\boldsymbol r-\boldsymbol d)}.
\end{equation}
Since first moments of the quadratures can be removed by local displacement operations, which do not affect separability, we focus below on centered Gaussian states, namely states with $\boldsymbol d=\boldsymbol 0$.

The purity of a Gaussian state is directly obtained from its covariance matrix. For a total of $2n$ modes, one has
\begin{equation}
    r_2=\Tr[\hat\rho^2]=\frac{1}{2^{2n}\sqrt{\det\gamma}}.
\end{equation}
We now turn to expression of the first-order weak moment $w_1$ for Gaussian states. Let
\begin{equation}
    \mathcal S_{\mathrm{BS}}=\frac{1}{\sqrt2}
    \begin{pmatrix}
    I_{2n} & I_{2n} \\
    -I_{2n} & I_{2n}
    \end{pmatrix}
\end{equation}
be the symplectic matrix describing $n$ balanced beam splitters, each mixing Alice's $j$th mode with Bob's $j$th mode. 
After this beam-splitter transformation, the covariance matrix becomes
\begin{equation}
    \gamma^{(1)}=\mathcal S_{\mathrm{BS}}\,\gamma\,\mathcal S_{\mathrm{BS}}^T.
\end{equation}
The measurement of $w_1$ consists, for each balanced beam splitter, in measuring the $p$ quadrature of the first output mode and the $x$ quadrature of the second output mode. We therefore introduce the restricted covariance matrix
\begin{equation}
    \gamma_{w_1}=\Pi_{w_1}\,\gamma^{(1)}\,\Pi_{w_1}^T,
\end{equation}
where $\Pi_{w_1}$ is the projection matrix that keeps only the measured quadratures
\begin{equation}
    (p_{A_1},\ldots,p_{A_n},x_{B_1},\ldots,x_{B_n})^T.
\end{equation}
at the output of the beam splitters. The joint probability density of obtaining zero outcomes for all these quadratures is
\begin{equation}
    P_{w_1}(\boldsymbol 0)=\frac{1}{(2\pi)^n\sqrt{\det\gamma_{w_1}}}.
\end{equation}
Using the relation between the maximally correlated vector $|\Omega\rangle$ and the homodyne projection discussed above, we obtain~\cite{Hertz2021}
\begin{equation}
    w_1=\pi^n P_{w_1}(\boldsymbol 0)=\frac{1}{2^n\sqrt{\det\gamma_{w_1}}}.
\end{equation}

The same reasoning applies to the second-order weak moment $w_2$, but with two copies of the Gaussian state. The covariance matrix of the two-copy state is
\begin{equation}
    \gamma\oplus\gamma.
\end{equation}
Let $\mathcal P_{B}^{(12)}$ denote the permutation matrix that exchanges Bob's modes between the two copies. This is the covariance-matrix representation of the swap $\hat S_{B_1B_2}$ appearing in the operator expression of $w_2$. After this swap, we apply balanced beam splitters to the pairs $A_1B_1$ and $A_2B_2$. At the covariance-matrix level, this gives
\begin{equation}
    \gamma^{(2)}=\mathcal S_{\mathrm{BS}}^{(2)}\mathcal P_{B}^{(12)}(\gamma\oplus\gamma)\left(\mathcal P_{B}^{(12)}\right)^T\left(\mathcal S_{\mathrm{BS}}^{(2)}\right)^T,
\end{equation}
where
\begin{equation}
    \mathcal S_{\mathrm{BS}}^{(2)}=\mathcal S_{\mathrm{BS}}\oplus \mathcal S_{\mathrm{BS}}.
\end{equation}

For both copies, the measured quadratures are the $p$ quadratures of the first output mode and the $x$ quadratures of the second output mode. We define the corresponding restricted covariance matrix as
\begin{equation}
    \gamma_{w_2}=\Pi_{w_2}\gamma^{(2)}\Pi_{w_2}^T,
\end{equation}
where $\Pi_{w_2}$ keeps the quadratures
\begin{equation}
\begin{aligned}
\Big(
&p^{(1)}_{A_1},\ldots,p^{(1)}_{A_n},
 x^{(1)}_{B_1},\ldots,x^{(1)}_{B_n},
\\
&p^{(2)}_{A_1},\ldots,p^{(2)}_{A_n},
 x^{(2)}_{B_1},\ldots,x^{(2)}_{B_n}
\Big)^T .
\end{aligned}
\end{equation}
The corresponding zero-outcome probability density is
\begin{equation}
    P_{w_2}(\boldsymbol 0)=\frac{1}{(2\pi)^{2n}\sqrt{\det\gamma_{w_2}}},
\end{equation}
and hence
\begin{equation}
    w_2=\pi^{2n}P_{w_2}(\boldsymbol 0)=\frac{1}{2^{2n}\sqrt{\det\gamma_{w_2}}}.
\end{equation}

Let us restrict the following discussion to a $1 \times 1$ Gaussian state. The covariance matrix can be brought by local Gaussian unitaries to the standard form~\cite{Weedbrook2012}
\begin{equation}
\label{eq:gaussian_covariance_matrix_abcd}
    \gamma=
    \begin{pmatrix}
    a & 0 & c & 0 \\
    0 & a & 0 & d \\
    c & 0 & b & 0 \\
    0 & d & 0 & b
    \end{pmatrix}.
\end{equation}
In this case, the above expressions reduce to simple closed formulas. The purity is
\begin{equation}
\label{eq:purity_gaussian_abcd}
    r_2=\frac{1}{4\sqrt{(ab-c^2)(ab-d^2)}}.
\end{equation}
For $w_1$, the restricted covariance matrix is
\begin{equation}
    \gamma_{w_1}=\frac12
    \begin{pmatrix}
    a+b+2d & 0\\
    0 & a+b-2c
    \end{pmatrix}.
\end{equation}
so that
\begin{equation}
\label{eq:w1_gaussian_abcd}
    w_1=\frac{1}{\sqrt{(a+b+2d)(a+b-2c)}}.
\end{equation}
For $w_2$, after swapping Bob's two copies and applying the two balanced beam splitters, the restricted covariance matrix is
\begin{equation}
    \gamma_{w_2}=\frac12
    \begin{pmatrix}
    a+b & 0 & 2d & 0\\
    0 & a+b & 0 & -2c\\
    2d & 0 & a+b & 0\\
    0 & -2c & 0 & a+b
    \end{pmatrix}.
\end{equation}
Therefore,
\begin{equation}
\label{eq:w2_gaussian_abcd}
    w_2=\frac{1}{\sqrt{\left[(a+b)^2-4c^2\right]\left[(a+b)^2-4d^2\right]}}.
\end{equation}

Let us now consider a locally phase-flipped noisy two-mode squeezed
vacuum (TMSV) state. It is related to the noisy TMSV by
a local phase shift of $\pi$ on Bob's mode and therefore has the same
entanglement (in the entropy of entanglement sense). Its covariance matrix is of the form of Eq.~\eqref{eq:gaussian_covariance_matrix_abcd}, with
\begin{equation}
a=b=\frac{\nu\cosh(2r)}{2},\quad c=-\frac{\nu\sinh(2r)}{2},\quad d=\frac{\nu\sinh(2r)}{2},
\end{equation}
where $r>0$ and $\nu\geq1$. It is pure when $\nu=1$, and mixed otherwise. Its entanglement is not detected by the weak realignment criterion of Eq.~\eqref{eq:weak_criterion}, but it can be detected by the second-moment criterion~\eqref{eq:our_criterion}.
Indeed, using Eq.~\eqref{eq:w1_gaussian_abcd}, we obtain
\begin{equation}
    w_1=\frac{e^{-2r}}{\nu}.
\end{equation}
Since $r>0$ and $\nu\geq 1$, one has $w_1\leq 1$. Hence, the weak realignment criterion does not detect entanglement for this family of states.
From Eq.~\eqref{eq:purity_gaussian_abcd}, the purity is $r_2 = {1}/{\nu^2}$. Similarly, using Eq.~\eqref{eq:w2_gaussian_abcd}, we obtain $w_2 = {1}/{\nu^2}$.
Thus, for all parameters satisfying
\begin{equation}
    1\leq \nu < \sqrt{2-e^{-4r}},
\end{equation}
the state is detected as entangled by the second-moment realignment criterion, while it remains undetected by the weak realignment criterion.

In particular, for pure TMSV states, corresponding to $\nu=1$, the second-moment realignment criterion gives
\begin{equation}
    r_2+\left(w_2-w_1^2\right) = 2-e^{-4r}.
\end{equation}
Since $r>0$, one has
\begin{equation}
    2-e^{-4r}>1,
\end{equation}
so every pure TMSV state with nonzero squeezing is certified as entangled.

\subsection{Non-Gaussian states}

\subsubsection{Mixed Schrödinger cat states}

We next consider a partially dephased two-mode Schrödinger cat state,
\begin{equation}
\label{eq:cat}
\begin{aligned}
\hat{\rho}_{\pm}
=\frac{1}{\mathcal{N}_{\pm}}\Big[
&\dyad{\alpha,\beta}
+\dyad{-\alpha,-\beta}
\\
&\pm(1-z)\Big(
|\alpha,\beta\rangle\langle-\alpha,-\beta|
+|-\alpha,-\beta\rangle\langle\alpha,\beta|
\Big)
\Big].
\end{aligned}
\end{equation}
where
\begin{equation}
    \mathcal{N}_{\pm}=2\left[1\pm(1-z)\exp^{-2\left(|\alpha|^2+|\beta|^2\right)}\right]
\end{equation}
ensures normalization. Here, $\alpha, \beta \in \mathbb C$, and the parameter $z\in[0,1]$ controls the coherence between the two product states $\ket{\alpha,\beta}$ and $\ket{-\alpha,-\beta}$.
For $z=0$, the coherence is maximal.
By contrast, for $z=1$, the coherence is minimal (equal to zero).
The $+$ and $-$ signs correspond to the even and odd cat states, respectively.

The first realignment moment of the state in Eq.~\eqref{eq:cat} evaluates to
\begin{equation}
\label{eq:cat-w1}
    w_1 =\frac{2\left[1\pm(1-z)\right]}{\mathcal{N}_{\pm}}e^{-|\alpha|^2-|\beta|^2+2\operatorname{Re}(\alpha\beta)}.
\end{equation}
For even cat states with real and positive $\alpha=\beta$, the weak criterion detects entanglement whenever $z<1$.
In contrast, for odd cat states, the weak realignment criterion cannot certify entanglement, since $w_1\leq 1$. This motivates the use of the second-moment criterion.
The second realignment moment is expressed as
\begin{equation}
\label{eq:cat-w2}
\begin{aligned}
w_2
={}&\frac{4e^{-2\left(|\alpha|^2+|\beta|^2\right)}}
{\mathcal{N}_{\pm}^{\,2}}
\Big\{
\left[1+(1-z)^2\right]
\cosh\!\left[4\operatorname{Re}(\alpha\beta)\right]
\\
&\pm 2(1-z)
\cos\!\left[4\operatorname{Im}(\alpha\beta)\right]
\Big\}.
\end{aligned}
\end{equation}
Finally, the purity, $r_2=\Tr\left[\hat{\rho}_{\pm}^{\,2}\right]$, is given by
\begin{equation}
\label{eq:cat-r2}
\begin{aligned}
r_2
={}&\frac{1}{\mathcal{N}_{\pm}^{\,2}}
\Big\{
2\left[1+e^{-4\left(|\alpha|^2+|\beta|^2\right)}\right]
\left[1+(1-z)^2\right]
\\
&\pm 8(1-z)
e^{-2\left(|\alpha|^2+|\beta|^2\right)}
\Big\}.
\end{aligned}
\end{equation}

\begin{figure}[h]
\includegraphics[width=0.8\columnwidth]{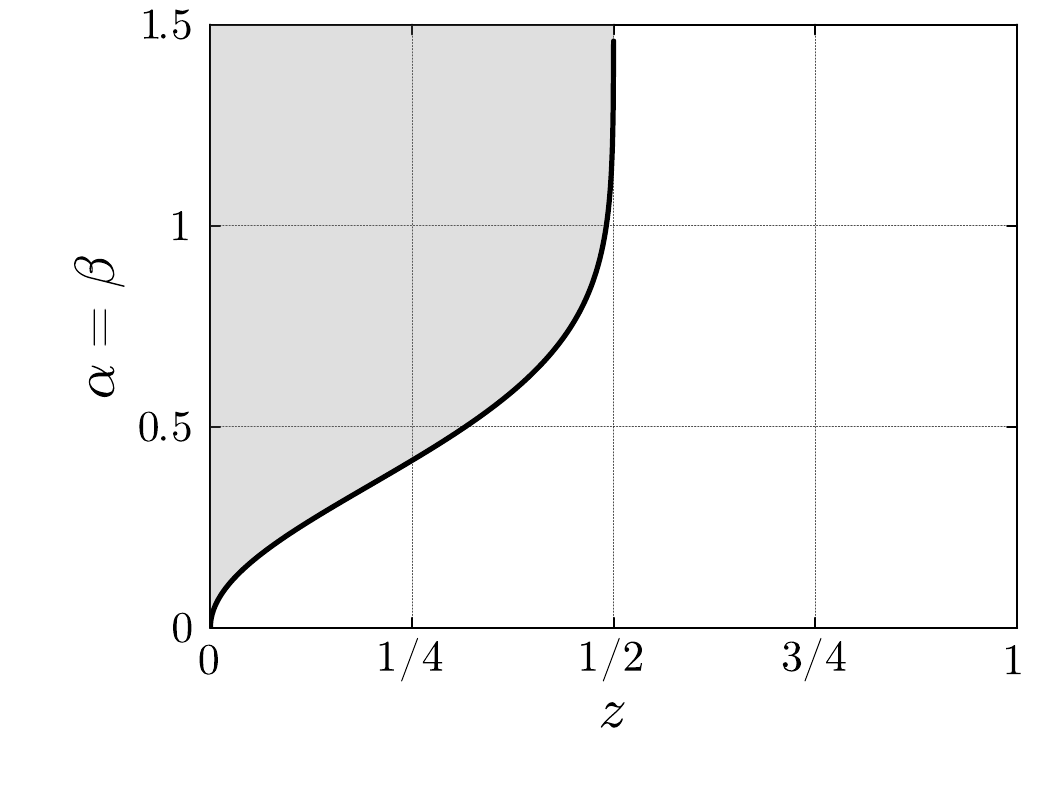}
    \caption{Detection boundary for odd cat states with real and positive parameters $\alpha=\beta$. The second-moment criterion detects entanglement above the curve $a(z)=\frac12\sqrt{-\ln(1-2z)}$, with $0\leq z<\frac12$. No violation occurs for $z\geq1/2$. For the even cat states, the weak criterion detects all states with $\alpha = \beta>0$ and $z<1$; the cases $\alpha = \beta =0$ or $z=1$ are separable.}
    \label{fig:odd-cat}
\end{figure}

These moments allow us to evaluate the measurable second-moment realignment criterion of Eq.~\eqref{eq:our_criterion} as displayed in Fig.~\ref{fig:odd-cat}.

\subsubsection{Lossy NOON states}

We now consider the experimentally motivated case of lossy NOON states. We start from the pure two-mode state (often simply called generalized NOON state)
\begin{equation}
    |\psi_N\rangle = \alpha |N,0\rangle+\beta |0,N\rangle, \qquad |\alpha|^2+|\beta|^2=1,
\end{equation}
and assume that the two modes of systems $A$ and $B$ undergo independent pure-loss channels $\mathcal L$ with transmissivities $\tau_A$ and $\tau_B$, respectively. 
A pure-loss channel of transmissivity $\tau$ can be modeled by coupling a mode of system $A$ to an environmental mode of system $E$ initially in the vacuum through a beam splitter of transmissivity $\tau$, namely,
\begin{equation}
\mathcal L_{\tau}(\hat\rho_A)
=
\Tr_E\left[
\hat U_{AE}^\tau
\left(\hat\rho_A\otimes|0\rangle\langle0|_E\right)
\left(\hat U_{AE}^\tau\right)^{\dagger}
\right],
\end{equation}
where $\hat U_{AE}^\tau$ is the corresponding beam-splitter unitary.

The resulting state is
\begin{equation}
    \hat\rho_N^{\mathrm{loss}}=\left(\mathcal{L}_{\tau_A} \otimes \mathcal{L}_{\tau_B}\right)\big(|\psi_N\rangle\langle\psi_N|\big).
\end{equation}
Explicitly, this gives
\begin{equation}
\begin{aligned}
\hat\rho_N^{\mathrm{loss}}
={}&(p_0+q_0)|0,0\rangle\langle 0,0|
\\
&+\sum_{k=1}^{N}p_k\,|k,0\rangle\langle k,0|
+\sum_{k=1}^{N}q_k\,|0,k\rangle\langle 0,k|
\\
&+\kappa\,|N,0\rangle\langle 0,N|
+\kappa^*\,|0,N\rangle\langle N,0|.
\end{aligned}
\end{equation}
where the different terms introduced are defined as
\begin{equation}
\begin{aligned}
p_k &\coloneqq
|\alpha|^2 \binom{N}{k}
\tau_A^k(1-\tau_A)^{N-k},
\\
q_k &\coloneqq
|\beta|^2 \binom{N}{k}
\tau_B^k(1-\tau_B)^{N-k},
\\
\kappa &\coloneqq
\alpha\beta^*(\tau_A\tau_B)^{N/2}.
\end{aligned}
\end{equation}
Applying the realignment map gives
\begin{equation}
\begin{aligned}
R(\hat\rho_N^{\mathrm{loss}})
={}&(p_0+q_0)|0,0\rangle\langle 0,0|
\\
&+\sum_{k=1}^{N}p_k\,|k,k\rangle\langle 0,0|+\sum_{k=1}^{N}q_k\,|0,0\rangle\langle k,k|
\\
&+\kappa\,|N,0\rangle\langle 0,N|
+\kappa^*\,|0,N\rangle\langle N,0|.
\end{aligned}
\end{equation}
Therefore,
\begin{equation}
    w_1=\Tr\left[R(\hat\rho_N^{\mathrm{loss}})\right]=p_0+q_0.
\end{equation}
Equivalently,
\begin{equation}
    w_1=|\alpha|^2(1-\tau_A)^N+|\beta|^2(1-\tau_B)^N.
\end{equation}
Since $0\leq \tau_A,\tau_B\leq 1$ and $|\alpha|^2+|\beta|^2=1$, one has \mbox{$w_1\leq 1$}. Hence, the weak realignment criterion never certifies entanglement for this family.

\begin{figure}[h]
    \centering
    \includegraphics[width=1.0\columnwidth]{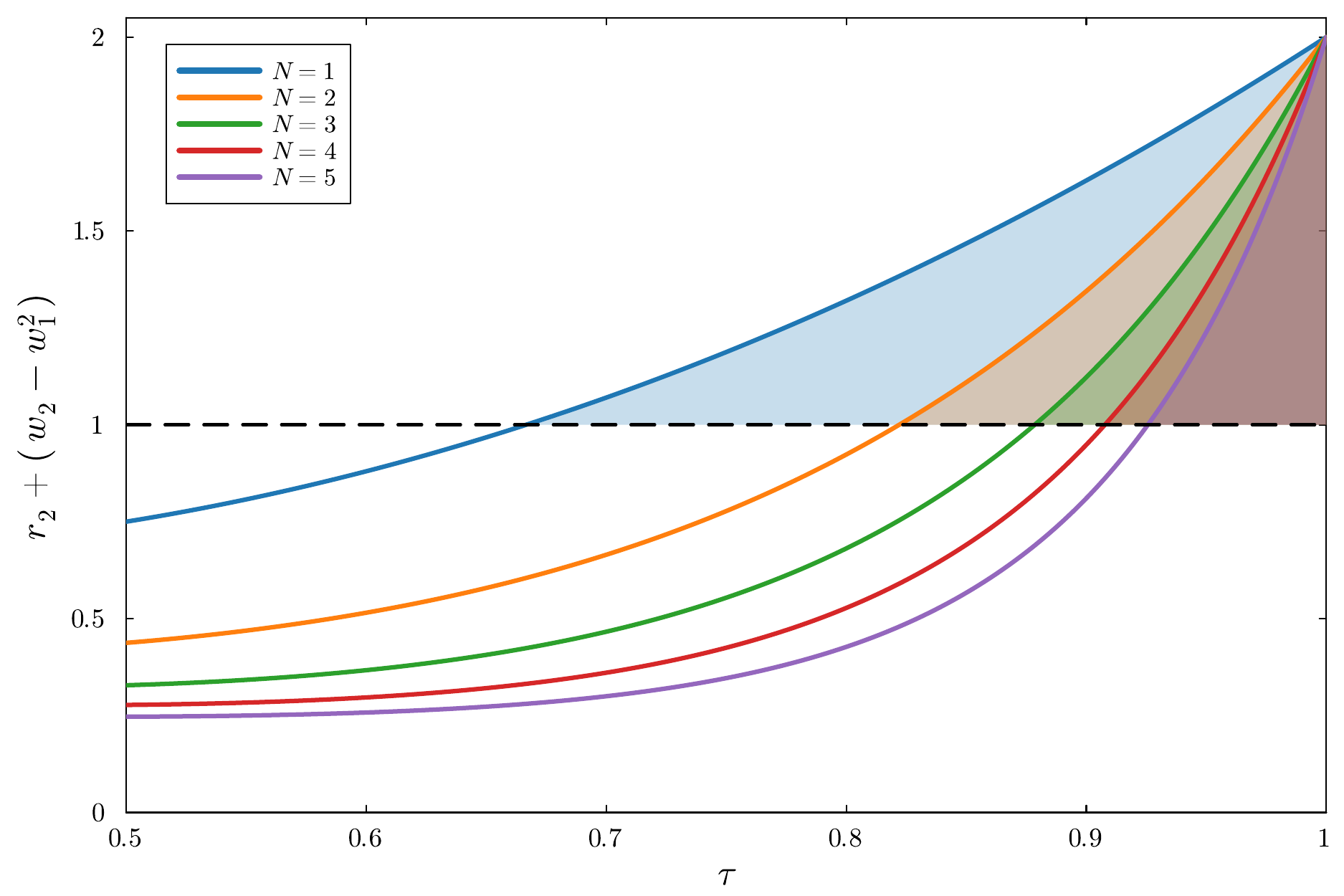}
    \caption{Value of the realignment criterion $r_2+(w_2-w_1^2)$ for balanced NOON states subjected to symmetric losses, $\tau_A=\tau_B=\tau$, for photon numbers $N=1,\ldots,5$. The horizontal dashed line indicates the separability bound equal to $1$. Entanglement is certified whenever a curve lies above this line. For each $N$, the criterion is therefore violated when the transmissivity exceeds a critical value, increasing with $N$.}
    \label{fig:lossy-noon}
\end{figure}

Let us now compute the quantities entering the improved criterion. The purity is
\begin{equation}
    r_2=(p_0+q_0)^2+\sum_{k=1}^{N}\left(p_k^2+q_k^2\right)+2|\kappa|^2.
\end{equation}
Similarly, the $w_2$ term is
\begin{equation}
    w_2=(p_0+q_0)^2+2\sum_{k=1}^{N}p_k q_k+2|\kappa|^2.
\end{equation}
For this family,
\begin{equation}
    w_2-w_1^2=2\sum_{k=1}^N p_kq_k+2|\kappa|^2\geq0.
\end{equation}
Hence, the absolute value in the criterion~\eqref{eq:our_criterion} reduces here to the plus branch,
\begin{equation}
    r_2+|w_2-w_1^2|=r_2+(w_2-w_1^2).
\end{equation}
Combining the above expressions for $w_1$, $w_2$, and $r_2$, we obtain
\begin{equation}
    r_2+\left(w_2-w_1^2\right)=\sum_{k=0}^{N}(p_k+q_k)^2+4|\kappa|^2.
\end{equation}
Therefore, the second-moment realignment criterion detects entanglement whenever
\begin{equation}
    \sum_{k=0}^{N}(p_k+q_k)^2+4|\kappa|^2>1.
\end{equation}
In Fig.~\ref{fig:lossy-noon}, we plot the value of the second-moment realignment criterion as a function of the transmissivity $\tau=\tau_A=\tau_B$ for balanced NOON states, i.e., $\alpha=\beta=1/\sqrt{2}$, and for photon numbers $N=1,\ldots,5$. For these states, the realigned operator is Hermitian, and hence $r_2=w_2$. The purity can therefore be inferred from the homodyne data used to estimate $w_2$.

In the lossless limit, when $\tau_A=\tau_B=1$, one has
\begin{equation}
    w_1=0,\qquad r_2=1,\qquad w_2=4|\alpha|^2|\beta|^2.
\end{equation}
Thus,
\begin{equation}
    r_2+\left(w_2-w_1^2\right)=1+4|\alpha|^2|\beta|^2.
\end{equation}
Hence, every entangled NOON state (every state with $\alpha\beta\neq 0$) is detected by the second-moment realignment criterion in the lossless case, whereas it remains undetected by the weak realignment criterion.

\section{Conclusion}
\label{sec:discussion}
We have derived a family of experimentally accessible separability criteria, which consist of inequalities involving both the singular moments $r_n$ and weak moments $w_n$ of the realigned density operator $R(\hat\rho)$. In contrast with previous investigations of realigned-moment separability criteria~\cite{Zhang2022,Wang2024Realignment,Aggarwal2024Realignment}, which were focused on finite-dimensional systems, our criteria are dimension-independent and hence experimentally tractable for CV quantum optical states.
Furthermore, our approach improves upon the weak realignment criterion~\cite{Hertz2021}, namely $w_1\le 1$, which can be interpreted as the first-order member of the family of realigned-moment criteria developed here. We have shown that the second-order member of this family,  namely $r_2 + |w_2 - w_1^2| \leq 1$ [Eq.~\eqref{eq:our_criterion}], is already sufficient to certify the entanglement of several classes of states that are left undetected by the weak realignment criterion, including mixed entangled Schrödinger cat states and NOON states.

Of course, a natural question is whether PPT entanglement might be detected by our second-moment realignment criterion~\eqref{eq:our_criterion}. This is a reasonable expectation since the latter lies between the original realignment criterion ($r_1\le 1$), which can detect PPT entanglement, and the weak realignment criterion ($w_1\le 1$), which is known to miss it. Unfortunately, upon finishing this work, it was brought to our attention that the second-moment criterion is provably weaker than the PPT criterion~\cite{MartinConversation}. Hence, it cannot detect bound entanglement, just as the weak realignment criterion. The proof is reported in Appendix~\ref{App:second-order-criterion+PPT}. Nevertheless, it remains unknown whether, in certain situations, higher-order realignment criteria may outperform the PPT criterion. This is conceivable, since the original realignment criterion does so, but we have not been able to establish that the family of realigned-moment criteria becomes monotonically stronger with increasing order. So far, we have not demonstrated that their detection capabilities surpass those of the separability criteria based on the moments of the partially transposed state studied in Ref.~\cite{Deside2026}. 

A potential advantage of the present approach might instead be on the experimental side because the weak moments $w_n$ (as well as even-order singular moments $r_{2n}$) of the realigned operator can be accessed using homodyne detection together with photon-number-parity measurements (which might even be achieved without recourse to photon-number counting~\cite{Plick2010Parity}). 
In contrast, the separability criteria based on the moments of the partially transposed state~\cite{Deside2026} require measuring the exact photon numbers. Specifically, our second-moment criterion involves only $w_1$, $w_2$, and $r_2$, all of them being measurable by homodyne and parity measurement with at most two copies of the state at a time, without requiring knowledge of the rank of the realigned operator (unlike the finite-dimensional criterion of, e.g., Ref.~\cite{Aggarwal2024Realignment}). Furthermore, if the realigned operator is known to be Hermitian, then $r_2=w_2$ so the purity $r_2$ can be obtained from the homodyne measurement of second weak moment $w_2$ without requiring any parity measurement at all.

Pushing further the comparison between our second-moment realignment criterion~\eqref{eq:our_criterion} and the criterion based on the third moment of the partially transposed state studied in Ref.~\cite{Deside2026}, we note that the total number of required copies 
is the same, namely five. However, the corresponding optical schemes are different. For the criterion based on partial transpose moments, the measurement of the third-order moment $p_3$ requires all three copies to be processed within a single interferometer, hence it requires a higher interferometric stability. (The last two copies must also be processed jointly to measure $p_2$, needed to apply the criterion.) In contrast, in the realignment-based protocol, the measurements of $w_1$ and $w_2$ can be performed separately, involving, respectively, one and two copies. The corresponding interferometric networks therefore require coherence between fewer copies at a time. (The last two copies must also be processed jointly to access $r_2$, needed to apply the criterion.)

Finally, an interesting direction for further work is to determine whether there exist physically relevant states that are not detected by the partial-transpose moment-based criteria but are detected by one or more realigned-moment criteria considered here. It would be worthwhile to identify specific regimes in which higher-order realignment moments provide a significant advantage over the second-moment criteria, and to investigate the robustness of the resulting protocols against realistic experimental imperfections.

\acknowledgments
We thank John Martin for providing us with a proof that our second-moment criterion \eqref{eq:our_criterion} cannot detect PPT entanglement. S.D. is a FRIA grantee of the Fonds de la Recherche Scientifique–FNRS (Belgium). N.J.C. and V.K. acknowledge support from the F.R.S.--FNRS under project CHEQS within the Excellence of Science (EOS) program. N.J.C. was further supported by the Fonds de la Recherche Scientifique–FNRS (Belgium) under Grant No.~T.0060.26.


\section*{Appendices}
\appendix

\section{Properties of the $n$th-order moments}
\label{app:properties of wn and rn}

Here, we prove several properties that $r_n$ and $w_n$ obey for all $n \in \mathbb N_{>0}$.
We first prove that
\begin{equation}
    r_n \coloneqq \Tr\left[\left(R(\hat{\rho})^\dagger R(\hat{\rho})\right)^{n/2}\right]
\end{equation}
is real and non-negative for all $n \in \mathbb N_{>0}$. To do so, we write the singular value decomposition of the realigned density matrix as
\begin{equation}
    R(\hat{\rho}) = U D V^\dagger,
\end{equation}
where $U$ and $V$ are unitary matrices, and $D$ is a diagonal matrix with non-negative entries $\{\lambda_i\}$. Then,
\begin{align}
    r_n&=\Tr\left[\left(V D^2 V^\dagger\right)^{n/2}\right] \\
    &=\Tr\left[\left(V D V^\dagger\right)^n\right] \\
    &=\Tr[D^n],
\end{align}
due to the unitarity of $U$ and $V$. Thus, the singular moment $r_n$ is simply the sum of the $n$th powers of the singular values of $R(\hat{\rho})$, that is, $r_n=\sum_i\lambda_i^n \geq0$.

We now turn to weak moments and prove that 
\begin{equation}
    w_n:=\Tr\left[R(\hat{\rho})^n\right]
\end{equation}
is real and non-negative for all $n \in \mathbb N_{>0}$.
Consider a pure-state decomposition of state $\hat\rho$,
\begin{equation}
    \hat\rho=\sum_\alpha p_\alpha
    \ket{\Psi_\alpha}\bra{\Psi_\alpha},
    \qquad p_\alpha\geq0.
\end{equation}
Using the vectorization of Eq.~\eqref{eq:ChoiJamiolkowski}, we associate with each state
\begin{equation}
    \ket{\Psi_\alpha}
    =\sum_{ij}(\Psi_\alpha)_{ij}\ket{i,j}
\end{equation}
an operator $\hat{\Psi}_\alpha$ such that 
\begin{equation}
    \hat\Psi_\alpha
    =\sum_{ij}(\Psi_\alpha)_{ij}\ket{i}\bra{j}.
\end{equation}
Using the definition of the realignment map $R(\hat A \otimes \hat B)=\ketbra{A}{B^*}$ and its dual expression $R(\ketbra{A}{B})=\hat A \otimes \hat B^*$, see~\cite{Hertz2021}, we find that $R\left(\ket{\psi_\alpha}\bra{\psi_\alpha}\right)
    =
    \hat\Psi_\alpha\otimes\hat\Psi_\alpha^*$.  Hence, 
\begin{equation}
    R(\hat\rho)
    =
    \sum_\alpha p_\alpha\,
    \hat\Psi_\alpha\otimes\hat\Psi_\alpha^*.
\end{equation}
Expanding the $n$th power of $R(\hat\rho)$ and taking the trace gives
\begin{align}
    w_n
    &=
    \sum_{\alpha_1,\ldots,\alpha_n}
    p_{\alpha_1}\cdots p_{\alpha_n}
    \left|
    \Tr\left[
    \hat\Psi_{\alpha_1}\cdots\hat\Psi_{\alpha_n}
    \right]
    \right|^2 ,
\end{align}
where we have used 
\begin{align}
\Tr\left[
    \hat\Psi_{\alpha_1}^*\cdots\hat\Psi_{\alpha_n}^*
    \right]
    =
    \Tr\left[
    \hat\Psi_{\alpha_1}\cdots\hat\Psi_{\alpha_n}
    \right]^*.
\end{align}
Thus, we have proven that $w_n \ge 0$.

We show next that
\begin{equation}
   w_n\leq r_n,
\end{equation}
for all $n\in\mathbb N_{>0}$.
Let ${x_i}$ denote the eigenvalues of $R(\hat{\rho})$, and let $\lambda_i$ denote its singular values. Then,
\begin{equation}
    w_n=\Tr\left[R(\hat{\rho})^n\right]=\sum_i x_i^n,
\end{equation}
whereas
\begin{equation}
    r_n=\sum_i \lambda_i^n.
\end{equation}
Using the triangle inequality and the standard (weak) majorization relation between eigenvalues and singular values~\cite{Marshall2011}, we obtain
\begin{equation}
    w_n\leq\sum_i|x_i|^n\leq\sum_i \lambda_i^n=r_n.
\end{equation}

Finally, we prove that $r_n\geq r_{n+1}$ for all $n\in\mathbb N_{>0}$. Since 
every singular value $\lambda_i$ of $R(\hat{\rho})$ satisfies $0\leq \lambda_i\leq1$, we have
\begin{equation}
    \lambda_i^n \geq \lambda_i^{n+1}
\end{equation}
for all $n\in \mathbb N_{>0}$, and summing over $i$ gives
\begin{equation}
    r_n\geq r_{n+1}.
\end{equation}

\section{Cross-Gram matrix from the operator Schmidt decomposition}
\label{app:A in B}
Here, we provide details on the properties of the cross-Gram matrix $M$ arising in the operator Schmidt decomposition of a mixed state, namely
\begin{equation}
    \hat\rho = \sum_i \lambda_i \, \hat A_i \otimes \hat B_i,
\end{equation}
where $\{\hat A_i\}$ and $\{\hat B_i\}$ are two orthonormal sets of operators with respect to the Hilbert--Schmidt inner product, i.e.,
\begin{equation}
\label{eq:orthoABs}
    \Tr\left[\hat A_i^\dagger \hat A_j\right]=\Tr\left[\hat B_i^\dagger \hat B_j\right]=\delta_{ij}.
\end{equation}
We assume that the local operator spaces are identified through a fixed basis, so that the operators $\hat A_i$ and $\hat B_i^*$ can be compared as elements of the same vector space.

We denote by $\{\hat B_j^*\}$ the set of entrywise complex conjugates of the operators $\{\hat B_j\}$ in the fixed basis. Since $\{\hat B_j^*\}$ is also an orthonormal set, each operator $\hat A_i$ can be decomposed into its projection onto the span of $\{\hat B_j^*\}$ and an orthogonal remainder, that is,
\begin{equation}
\label{eq:AinBs}
    \hat A_i=\sum_j m_j^{(i)}\,\hat B_j^*+\hat C_i,
\end{equation}
where $ m_j^{(i)} \in \mathbb C$ and $\hat C_i$ is orthogonal to the span of $\{\hat B_j^*\}$, namely,
\begin{equation}
    \Tr\left[\left(\hat B_j^*\right)^\dagger \hat C_i\right]=\Tr\left[\hat B_j^T \hat C_i\right]=0\qquad\forall i,j.
\end{equation}
Using the orthonormality of the operators in the sets $\{\hat A_i\}$ and $\{\hat B_i^*\}$, see Eq.~\eqref{eq:orthoABs}, and Eq.~\eqref{eq:AinBs}, we get
\begin{equation}
    \sum_k\left(m_k^{(i)}\right)^*m_k^{(j)}+\Tr\left[\hat C_i^\dagger \hat C_j\right]=\delta_{ij}.
\end{equation}
Introducing the matrix $M$ with entries
\begin{equation}
\label{eq:Mandm}
    M_{ij}=m_i^{(j)},
\end{equation}
and the Gram matrix $N_C$ of the residual operators,
\begin{equation}
    \left(N_C\right)_{ij}=\Tr\left[\hat C_i^\dagger \hat C_j\right],
\end{equation}
we may write
\begin{equation}
    M^\dagger M+N_C=I.
\end{equation}
Since $N_C$ is positive semidefinite, it follows that
\begin{equation}
  \|M\|_2 \equiv   M^\dagger M\leq I,
\end{equation}
where $\|\cdot\|_2$ denotes the spectral norm. Thus, the matrix $M$ is a contraction.

As we deduce from Eq.~\eqref{eq:Mandm}, the coefficients appearing in Eq.~\eqref{eq:AinBs} are the Hilbert--Schmidt overlaps between $\hat A_j$ and $\hat B_i^*$. Hence,
\begin{equation}
    M_{ij}=\Tr\left[\left(\hat B_i^*\right)^\dagger \hat A_j\right]=\Tr\left[\hat B_i^T \hat A_j\right].
\end{equation}

\section{Proof of $|M^{[ij]}|\le 1$}

\label{app:w2bound}

In order to derive the second-moment realignment criterion, we need to prove that
\begin{equation}
    \left|M_{ii}M_{jj}-M_{ij}M_{ji}\right|\leq 1.
\end{equation}
We use the result of Appendix~\ref{app:A in B}, namely the fact that $M$ is a contraction.
Let $M^{[ij]}$ denote the $2\times 2$ submatrix of $M$ obtained by keeping rows $i,j$ and columns $i,j$, namely,
\begin{equation}
    M^{[ij]}=
    \begin{pmatrix}
    M_{ii} & M_{ij}\\
    M_{ji} & M_{jj}
    \end{pmatrix}.
\end{equation}
Equivalently, one may write
\begin{equation}
    M^{[ij]}=P^{[ij]} M Q^{[ij]},
\end{equation}
where $P^{[ij]}$ is the $2\times r$ matrix selecting rows $i,j$, and $Q^{[ij]}$ is the $r\times 2$ matrix selecting columns $i,j$, with $r$ being the Schmidt rank. Both are partial isometries, so
\begin{equation}
    \left\|P^{[ij]}\right\|_2=\left\|Q^{[ij]}\right\|_2=1.
\end{equation}
Then, by submultiplicativity of the spectral norm and the fact that $M$ is a contraction,
\begin{equation}
    \left\|M^{[ij]}\right\|_2 \leq\left\|P^{[ij]}\right\|_2\big\|M\big\|_2\left\|Q^{[ij]}\right\|_2\leq1.
\end{equation}
Let $s_1$ and $s_2$ be the singular values of $M^{[ij]}$. Since
\begin{equation}
    \left\|M^{[ij]}\right\|_2=\max\{s_1,s_2\}\leq 1,
\end{equation}
we have $s_1\leq 1$ and $s_2\leq 1$. Therefore,
\begin{equation}
    \left|\det M^{[ij]} \right| = s_1s_2 \leq 1,
\end{equation}
implying that
\begin{equation}
   \left|\det M^{[ij]}\right|= \left|M_{ii}M_{jj}-M_{ij}M_{ji}\right|\leq 1.
\end{equation}

In Appendix~\ref{app:more than 2}, we generalize this idea to higher-order determinants, which leads to higher-order realignment criteria.

\section{Higher-order realignment-moment criteria} 
\label{app:more than 2}
We extend here the argument of Appendix~\ref{app:w2bound} to higher-order minors of the matrix $M$. 
Recall that the realigned operator can be written as (see Eq.~\eqref{eq:realigned_schmidt})
\begin{equation} 
    R(\hat\rho) = \sum_i \lambda_i |A_i\rangle\langle B_i^*|, 
\end{equation} 
where $\lambda_i\geq 0$, and define matrix $M$ by 
\begin{equation} 
    M_{ij} = \Tr\left[ \hat B_i^T \hat A_j \right]. 
\end{equation} 
From Appendix~\ref{app:A in B}, we have that $M$ is a contraction, i.e.,
\begin{equation} 
    M^\dagger M\leq I. 
\end{equation}
We introduce the moments 
\begin{equation} 
    r_n = \sum_i \lambda_i^n, 
\end{equation} 
and 
\begin{equation} 
    w_n = \sum_{i_1,\ldots,i_n} \lambda_{i_1}\cdots \lambda_{i_n} M_{i_1 i_2}M_{i_2 i_3}\cdots M_{i_n i_1},
\end{equation} 
and refer to Appendix~\ref{app:general-multicopy} for the quantum-optical measurements of $w_n$ and $r_{2n}$.

Let us first consider the third-order case. For three distinct indices $i,j,k$, define the $3\times 3$ principal submatrix 
\begin{equation} 
    M^{[ijk]} = 
    \begin{pmatrix} 
    M_{ii} & M_{ij} & M_{ik}\\ 
    M_{ji} & M_{jj} & M_{jk}\\ 
    M_{ki} & M_{kj} & M_{kk} 
    \end{pmatrix}. 
\end{equation} 
Its determinant is 
\begin{equation} 
    \begin{aligned} 
    \det M^{[ijk]} = & M_{ii}M_{jj}M_{kk} + M_{ij}M_{jk}M_{ki} + M_{ik}M_{ji}M_{kj} \\ 
    & - M_{ik}M_{jj}M_{ki} - M_{ij}M_{ji}M_{kk} - M_{ii}M_{jk}M_{kj}. 
    \end{aligned} 
\end{equation} 
Since $M$ is a contraction, every principal submatrix is also a contraction. Therefore, 
\begin{equation} 
    \left\| M^{[ijk]} \right\|_2 \leq 1. 
\end{equation} 
If $s_1,s_2,s_3$ are the singular values of $M^{[ijk]}$, then $s_\alpha\leq 1$ for all $\alpha$, and hence 
\begin{equation} 
    \left| \det M^{[ijk]} \right| = s_1s_2s_3 \leq 1. 
\end{equation} 
We now use this bound to derive a third-order inequality. Notice that 
\begin{equation} 
    r_1^3-3r_1r_2+2r_3 = \sum_{\substack{i,j,k\\ \text{all distinct}}} \lambda_i\lambda_j\lambda_k. 
\end{equation} 
Similarly, expanding the weak moments gives 
\begin{equation} 
    \begin{aligned} 
    w_1^3-3w_1w_2+2w_3 = \sum_{\substack{i,j,k\\ \text{all distinct}}} \lambda_i\lambda_j\lambda_k \det M^{[ijk]}. 
    \end{aligned} 
\end{equation} 
Consequently, 
\begin{equation}
\label{eq:third-order-ineq}
\begin{aligned}
&r_1^3-3r_1r_2+2r_3
\pm \left(w_1^3-3w_1w_2+2w_3\right)
\\
&\qquad=
\sum_{\substack{i,j,k\\ \text{all distinct}}}
\lambda_i\lambda_j\lambda_k
\left(1\pm\det M^{[ijk]}\right).
\end{aligned}
\end{equation}
Although the right-hand side of Eq.~\eqref{eq:third-order-ineq} is real after summation, the individual determinants $\det M^{[ijk]}$ need not be real. Taking the real part term by term gives
\begin{equation}
\begin{aligned}
&r_1^3-3r_1r_2+2r_3
\pm\left(w_1^3-3w_1w_2+2w_3\right)
\\
&\qquad=
\sum_{\substack{i,j,k\\ \mathrm{all\ distinct}}}
\lambda_i\lambda_j\lambda_k
\left(1\pm
\mathrm{Re}\!\left[\det M^{[ijk]}\right]\right).
\end{aligned}
\end{equation}
Since
\begin{equation}
    \left|\mathrm{Re}\!\left(\det M^{[ijk]}\right)\right|\leq\left|\det M^{[ijk]}\right|\leq1,
\end{equation}
each term in the last sum is non-negative. Hence, we obtain 
\begin{equation} 
    r_1^3-3r_1r_2+2r_3 \pm (w_1^3-3w_1w_2+2w_3) \geq 0, 
\end{equation}
or equivalently, 
\begin{equation} 
    3r_1r_2-2r_3\mp\left(w_1^3-3w_1w_2+2w_3\right) \leq r_1^3. 
\end{equation} 
For separable states, the realignment criterion $r_1\leq1$ implies $r_1^3\leq 1$, so the previous inequalities yields the following \textit{third-moment realignment criterion}
\begin{equation}
\hat{\rho}\ \mathrm{ separ.}\implies    3r_1r_2-2r_3+
    \left|w_1^3-3w_1w_2+2w_3\right|\leq1.
    \label{eq:third-order-unrelaxed}
\end{equation}
Note that the quantities $r_1$ and $r_3$ are not directly accessible with the multicopy measurement scheme considered in this work because these are odd-order $r_n$-moments (see Appendix~\ref{app:general-multicopy}). However, since $r_1\geq w_1$ and $r_3\leq r_2$, we may write a relaxed (weaker but experimentally accessible) bound for separable states,
\begin{equation}
\label{eq:third-moment-realignment}
    3w_1r_2-2r_2+\left|w_1^3-3w_1w_2+2w_3\right|\leq1.
\end{equation} 
Hence, any measured violation to inequality~\eqref{eq:third-moment-realignment} certifies entanglement. 

We can now state the generalization\footnote{The connection with elementary symmetric polynomials was suggested by GPT-5.6 Sol. The subsequent development of the generalization was carried out with assistance from GPT-5.6 Sol and was independently verified in full.} to order $\ell$. Let
\begin{equation}
    I=\{i_1,\ldots,i_\ell\}
\end{equation}
be a set of $\ell$ distinct indices, and let $M^{[I]}$ denote the $\ell\times \ell$ principal submatrix of $M$ obtained by keeping the rows and columns indexed by $I$. Since $M$ is a contraction, every such principal submatrix is also a contraction. Hence,
\begin{equation}
    \left\|M^{[I]}\right\|_2\leq 1.
\end{equation}
If $s_1,\ldots,s_\ell$ are the singular values of $M^{[I]}$, then $s_\alpha\leq1$ for all $\alpha$, and therefore
\begin{equation}
    \left|\det M^{[I]}\right|=\prod_{\alpha=1}^{\ell} s_\alpha\leq 1.
\end{equation}

Since determinants of $\ell\times \ell$ submatrices are associated with products of $\ell$ distinct Schmidt coefficients,
\begin{equation}
    \prod_{i\in I}\lambda_i,
\end{equation}
summing over all subsets $I$ of cardinality $\ell$ gives the $\ell$th elementary symmetric polynomial~\cite{Macdonald1995SymmetricFunctions},
\begin{equation}
    e_\ell(\lambda)\coloneqq\sum_{\substack{I\\ |I|=\ell}}\prod_{i\in I}\lambda_i .
\end{equation}
However, the quantities we use in the construction of the criteria are the moments
\begin{equation}
    r_k=\sum_i \lambda_i^k.
\end{equation}
These are the power sums of the Schmidt coefficients. Newton's identities allow one to express the elementary symmetric polynomials in terms of these power sums~\cite{Macdonald1995SymmetricFunctions}. For this reason, we define
\begin{equation}
    \mathcal P_\ell(x_1,\ldots,x_\ell)\coloneqq\ell!\,e_\ell,
\end{equation}
where $e_\ell$ is expressed in terms of the power sums $x_1,\ldots,x_\ell$.

Equivalently,
\begin{equation}
    \mathcal P_\ell(x_1,\ldots,x_\ell)=\det
    \begin{pmatrix}
    x_1 & 1 & 0 & \cdots & 0\\
    x_2 & x_1 & 2 & \cdots & 0\\
    x_3 & x_2 & x_1 & \ddots & \vdots\\
    \vdots & \vdots & \vdots & \ddots & \ell-1\\
    x_\ell & x_{\ell-1} & x_{\ell-2} & \cdots & x_1
    \end{pmatrix}.
\end{equation}
Thus, for the Schmidt coefficients,
\begin{equation}
    \mathcal P_\ell(r_1,\ldots,r_\ell)=\ell!\sum_{\substack{I\\ |I|=\ell}}\prod_{i\in I}\lambda_i.
\end{equation}
For example, at order $2$,
\begin{equation}
    \mathcal P_2(x_1,x_2)=x_1^2-x_2,
\end{equation}
so that
\begin{equation}
    \mathcal P_2(r_1,r_2)=r_1^2-r_2=\sum_{i\neq j}\lambda_i\lambda_j .
\end{equation}
At order $3$,
\begin{equation}
    \mathcal P_3(x_1,x_2,x_3)=x_1^3-3x_1x_2+2x_3,
\end{equation}
and therefore
\begin{equation}
    \mathcal P_3(r_1,r_2,r_3)=r_1^3-3r_1r_2+2r_3=\sum_{\substack{i,j,k\\ \mathrm{all\ distinct}}}\lambda_i\lambda_j\lambda_k.
\end{equation}
This is precisely the sum over products of three distinct Schmidt coefficients.

The same structure appears for the weak moments,
\begin{equation}
    w_n\coloneqq\Tr\left[R(\hat\rho)^n\right],
\end{equation}
which are the power sums of the eigenvalues of $R(\hat\rho)$. Applying Newton's identities to these moments gives
\begin{equation}
    \mathcal P_\ell(w_1,\ldots,w_\ell)=\ell!\sum_{\substack{I\\ |I|=\ell}}\left(\prod_{i\in I}\lambda_i\det M^{[I]}\right).
\end{equation}
Combining the two expressions, we obtain
\begin{equation}
\begin{aligned}
&\mathcal P_\ell(r_1,\ldots,r_\ell)
\pm \mathcal P_\ell(w_1,\ldots,w_\ell)
\\
&\qquad=
\ell!\sum_{\substack{I\\ |I|=\ell}}
\left(\prod_{i\in I}\lambda_i\right)
\left(
1\pm\mathrm{Re}\!\left[\det M^{[I]}\right]
\right).
\end{aligned}
\end{equation}
Since $\left|\det M^{[I]}\right|\leq1$ and the moments $w_n$ are real (see Appendix~\ref{app:properties of wn and rn}), this reduces to
\begin{equation}
    \mathcal P_\ell(r_1,\ldots,r_\ell)\pm\mathcal P_\ell(w_1,\ldots,w_\ell)\geq0.
\end{equation}

Thus, each value of $\ell$ yields an inequality involving the moments up to order $\ell$ that follows from the contraction property $M^\dagger M\leq I$. Together with the separability condition $r_1\leq1$, these inequalities generate a family of higher-order realignment-moment separability criteria. 
To obtain a directly measurable separability criterion at a given order $\ell$ , the odd moments $r_{2n+1}$ must additionally be eliminated or bounded using the experimentally accessible quantities $w_n$ and $r_{2n}$ (see Appendix~\ref{app:general-multicopy} for the quantum-optical measurement methods). The third-moment criterion derived in Eq.~\eqref{eq:third-order-unrelaxed} and its relaxed version, Eq.~\eqref{eq:third-moment-realignment}, provide an explicit example of such a reduction.

\section{General measurement techniques for $w_n$ and $r_{2n}$}
\label{app:general-multicopy}

Here, we derive multicopy observables for the weak moments $w_n$ and the even moments $r_{2n}$, and describe their interferometric implementation. Recall that
\begin{equation}
    w_n=\Tr\left[R(\hat\rho)^n\right],
\end{equation}
whereas
\begin{equation}
    r_{2n}=\Tr\left[\left(R(\hat\rho)^\dagger R(\hat\rho)\right)^n\right].
\end{equation}
The quantity $w_n$ is a homogeneous polynomial of degree $n$ in the matrix elements of $\hat\rho$, while $r_{2n}$ is a homogeneous polynomial of degree $2n$. They can therefore be accessed using $n$ and $2n$ copies of the state, respectively~\cite{Brun2004}.

The multicopy method~\cite{Brun2004} allows one to access any homogeneous polynomial function of the matrix elements of a density operator. Consider an arbitrary bipartite density operator
\begin{equation}
\label{app:eq:density_operator}
    \hat\rho=\sum_{ijkl}\rho_{ij,kl}\ket{i,j}\bra{k,l},
\end{equation}
where $\rho_{ij,kl} = \langle i, j| \hat \rho| k,l\rangle \in \mathbb C$.
A homogeneous polynomial of degree $n$ can be written as
\begin{equation}
    f_n(\hat\rho)=\sum_{\substack{i_1j_1k_1l_1\\\vdots\\i_nj_nk_nl_n}}c_{i_1j_1k_1l_1\cdots i_nj_nk_nl_n}\rho_{i_1j_1,k_1l_1}\cdots\rho_{i_nj_n,k_nl_n},
\end{equation}
where the coefficients $c_{i_1j_1k_1l_1\cdots i_nj_nk_nl_n}$ are independent of $\hat\rho$. Such a polynomial can be expressed as
\begin{equation}
    f_n(\hat\rho)=\Tr\left[\hat\rho^{\otimes n}\hat O_n\right]
\end{equation}
for an appropriate multicopy operator $\hat O_n$.

In order to express the multicopy operators corresponding to $w_n$ and $r_{2n}$, we begin by introducing the right cyclic-shift operator acting on $n$ copies,
\begin{equation}
    \hat{\overrightarrow{\Pi}}_n\ket{i_1,i_2,\ldots,i_n}=\ket{i_n,i_1,\ldots,i_{n-1}},
\end{equation}
and the corresponding left cyclic-shift operator,
\begin{equation}
    \hat{\overleftarrow{\Pi}}_n=\left(\hat{\overrightarrow{\Pi}}_n\right)^\dagger,
\end{equation}
which acts as
\begin{equation}
    \hat{\overleftarrow{\Pi}}_n\ket{i_1,i_2,\ldots,i_n}=\ket{i_2,\ldots,i_n,i_1}.
\end{equation}
In what follows, the superscripts $A$ and $B$ indicate whether the shift acts on Alice's or Bob's copies.

\subsection{Multicopy observable for $w_n$}
\label{app:multicopy_observable_for_wn}

Using the density operator as expressed in Eq.~\eqref{app:eq:density_operator}, the trace of the $n$th power of the realigned operator reads
\begin{equation}
    w_n=\sum_{\substack{i_1,\ldots,i_n\\k_1,\ldots,k_n}}\rho_{i_1i_2,k_1k_2}\, \rho_{i_2i_3,k_2k_3}\cdots\,\rho_{i_ni_1,k_nk_1}.
\end{equation}
Then, since
$\Tr\left[\hat\rho \ket{k,l}\bra{i,j}\right]=\rho_{ij,kl}$,
we obtain
\begin{equation}
\label{eq:multicopy-wn}
    w_n=\Tr\left[\hat\rho^{\otimes n}\, \hat O_n^{(w)}\right],
\end{equation}
where
\begin{equation}
\label{app:eq:multicopy_for_wn}
\begin{aligned}
    \hat O_n^{(w)}=\sum_{\substack{i_1,\ldots,i_n\\k_1,\ldots,k_n}}\bigotimes_{t=1}^{n}\ket{k_t,k_{t+1}}\bra{i_t,i_{t+1}},
\end{aligned}
\end{equation}
where $i_{n+1}\coloneqq i_1$ and $k_{n+1}\coloneqq k_1$.
To express this operator in terms of cyclic shifts, we use
\begin{equation}
    R\left(\ket{k_t,i_t}\bra{k_{t+1},i_{t+1}}\right)=\ket{k_t,k_{t+1}}\bra{i_t,i_{t+1}}.
\end{equation}
It follows that
\begin{equation}
    \hat O_n^{(w)}=R^{\otimes n}\left[\sum_{\substack{i_1,\ldots,i_n\\k_1,\ldots,k_n}}\bigotimes_{t=1}^{n}\ket{k_t,i_t}\bra{k_{t+1},i_{t+1}}\right].
\end{equation}
After reordering the tensor factors from $(A_1B_1) \cdots (A_nB_n)$ to $(A_1\cdots A_n) (B_1\cdots B_n)$, the operator inside the brackets factorizes as
\begin{equation}
    \sum_{\substack{i_1,\ldots,i_n\\k_1,\ldots,k_n}}\bigotimes_{t=1}^{n}\ket{k_t,i_t}\bra{k_{t+1},i_{t+1}}=\hat{\overrightarrow{\Pi}}_n^A\otimes\hat{\overrightarrow{\Pi}}_n^B.
\end{equation}
Consequently,
\begin{equation}
\label{eq:observable_wn_as_cyclic_shifts}
    \hat O_n^{(w)}=R^{\otimes n}\left(\hat{\overrightarrow{\Pi}}_n^A\otimes\hat{\overrightarrow{\Pi}}_n^B\right),
\end{equation}
and we have
\begin{equation}
\label{eq-new-expression-of-w_n}
w_n= \tr \left[ \hat\rho^{\otimes n} \, R^{\otimes n}\left(\hat{\overrightarrow{\Pi}}_n^A\otimes\hat{\overrightarrow{\Pi}}_n^B\right) \right] .
\end{equation}
Note that this expression can be directly understood by applying the multicopy technique on  $R(\hat\rho)^{n}$, that is
\begin{equation}
w_n= \tr \left[ R(\hat\rho)^n \right] = \tr \left[ R(\hat\rho)^{\otimes n} \left(\hat{\overrightarrow{\Pi}}_n^A\otimes\hat{\overrightarrow{\Pi}}_n^B\right) \right]   .
\end{equation}
Indeed, since the dual of the realignment map is equal to itself \cite{Hertz2021}, we may apply $R^{\otimes n}$ to the second factor instead of the first one inside the trace, yielding Eq.~\eqref{eq-new-expression-of-w_n}.

In order to find a quantum-optical measurement setting for $w_n$, it is also useful to express this operator directly in terms of EPR projectors. Using the unnormalized EPR-like vector
\begin{equation}
    \ket{\Omega}\coloneqq\sum_{q=0}^{\infty}\ket{q,q},
\end{equation}
we can regroup the factors of $\hat O_n^{(w)}$ (see Eq.~\eqref{app:eq:multicopy_for_wn}) according to the pairs $(A_t,B_{t-1})$, with the cyclic convention $B_{0}\coloneqq B_n$, to write
\begin{equation}
\label{eq:observable-wn-epr}
    \hat O_n^{(w)}=\bigotimes_{t=1}^{n}\ket{\Omega}\!\bra{\Omega}_{A_tB_{t-1}}.
\end{equation}
Therefore,
\begin{equation}
\label{eq:wn-epr-projectors}
    w_n=\Tr\left[\hat\rho^{\otimes n}\bigotimes_{t=1}^{n}\ket{\Omega}\!\bra{\Omega}_{A_tB_{t-1}}\right].
\end{equation}
Equation~\eqref{eq:observable-wn-epr} also confirms directly that $\hat O_n^{(w)}$ is Hermitian and positive semidefinite, and hence that $w_n$ is real and non-negative.

It is instructive to compare the expression of the multicopy observable in Eq.~\eqref{eq:observable_wn_as_cyclic_shifts} with the multicopy expression for the moments of the partially transposed state~\cite{Elben2020,Deside2026}, namely
\begin{equation}
\begin{aligned}
    p_n&\coloneqq\Tr\left[\left(\hat\rho^{T_B}\right)^n\right]\\
    &=\Tr\left[\hat\rho^{\otimes n}\left(\hat{\overrightarrow{\Pi}}_n^A\otimes\hat{\overleftarrow{\Pi}}_n^B\right)\right],
\end{aligned}
\end{equation}
where $T_B$ denotes the partial transposition on Bob's side.
For the partial-transpose moments, the cyclic shifts on Alice's and Bob's copies have opposite orientations. By contrast, the expression for $w_n$ involves shifts in the same direction, followed by the independent application of the realignment map on each copy.

\subsection{Multicopy observable for $r_{2n}$}

Using the density operator of Eq.~\eqref{app:eq:density_operator}, we may write
\begin{equation}
\begin{aligned}
    r_{2n}=\sum_{\substack{i_1,\ldots,i_n\\j_1,\ldots,j_n\\k_1,\ldots,k_n\\l_1,\ldots,l_n}}\prod_{t=1}^{n}\rho_{i_t j_t, k_t l_t}^*\rho_{i_t j_{t+1}, k_t l_{t+1}},
\end{aligned}
\end{equation}
where $j_{n+1}\coloneqq j_1$ and $l_{n+1}\coloneqq l_1$.
Then, using the Hermiticity of $\hat\rho$, i.e., $\rho_{i_tj_t,k_tl_t}^*=\rho_{k_tl_t,i_tj_t}$, we obtain
\begin{equation}
    r_{2n}=\sum_{\substack{i_1,\ldots,i_n\\j_1,\ldots,j_n\\k_1,\ldots,k_n\\l_1,\ldots,l_n}}\prod_{t=1}^{n}\rho_{k_tl_t,i_tj_t}\rho_{i_tj_{t+1},k_tl_{t+1}}.
\end{equation}

Each term contains $2n$ matrix elements of $\hat\rho$, so the moment can be written as an expectation value on $2n$ copies, that is,
\begin{equation}
\label{eq:multicopy-r2n}
    r_{2n}=\Tr\left[\hat\rho^{\otimes2n}\hat O_{2n}^{(r)}\right],
\end{equation}
where
\begin{equation}
\begin{aligned}
\hat O_{2n}^{(r)}
={}&
\sum_{\substack{
i_1,\ldots,i_n;\,j_1,\ldots,j_n\\
k_1,\ldots,k_n;\,l_1,\ldots,l_n
}}
\bigotimes_{t=1}^{n}
\Big[
\left(\ket{i_t,j_t}\bra{k_t,l_t}\right)_{2t-1}
\\
&\qquad\qquad\otimes
\left(\ket{k_t,l_{t+1}}\bra{i_t,j_{t+1}}\right)_{2t}
\Big].
\end{aligned}
\end{equation}

After regrouping the tensor factors from $(A_1B_1)\cdots (A_{2n}B_{2n})$ to $(A_1\cdots A_{2n}) (B_1\cdots B_{2n})$, the operator factorizes into independent permutations on Alice's and Bob's subsystems. Indeed, on Alice's side,
\begin{equation}
\begin{aligned}
&\sum_{\substack{i_1,\ldots,i_n\\ k_1,\ldots,k_n}}
\bigotimes_{t=1}^{n}
\Big[
\left(\ket{i_t}\bra{k_t}\right)_{A_{2t-1}}
\otimes
\left(\ket{k_t}\bra{i_t}\right)_{A_{2t}}
\Big]
\\
&\qquad=
\bigotimes_{t=1}^{n}
\hat S_{A_{2t-1},A_{2t}} .
\end{aligned}
\end{equation}
where $\hat S_{X,Y}$ denotes the swap operator between subsystems $X$ and $Y$, i.e.,
\begin{equation}
    \hat S_{X,Y}\ket{u}_X\ket{v}_Y=\ket{v}_X\ket{u}_Y.
\end{equation}
Similarly, on Bob's side,
\begin{equation}
\begin{aligned}
&\sum_{\substack{j_1,\ldots,j_n\\ l_1,\ldots,l_n}}
\bigotimes_{t=1}^{n}
\Big[
\left(\ket{j_t}\bra{l_t}\right)_{B_{2t-1}}
\otimes
\left(\ket{l_{t+1}}\bra{j_{t+1}}\right)_{B_{2t}}
\Big]
\\
&\qquad=
\bigotimes_{t=1}^{n}
\hat S_{B_{2t},B_{2t+1}} .
\end{aligned}
\end{equation}
with the convention $B_{2n+1}\coloneqq B_1$.
Thus, combining the two sides gives
\begin{equation}
\label{eq:observable-r2n}
    \hat O_{2n}^{(r)}=\left(\bigotimes_{t=1}^{n}\hat S_{A_{2t-1},A_{2t}}\right)\otimes\left(\bigotimes_{t=1}^{n}\hat S_{B_{2t},B_{2t+1}}\right).
\end{equation}

In summary, Alice's subsystems are paired according to
\begin{equation}
    (A_1,A_2),\,(A_3,A_4),\,\ldots,\,(A_{2n-1},A_{2n}),
\end{equation}
whereas Bob's subsystems follow the staggered cyclic pairing
\begin{equation}
    (B_2,B_3),\,(B_4,B_5),\,\ldots,\,(B_{2n},B_1).
\end{equation}
The operator $\hat O_{2n}^{(r)}$ is Hermitian and unitary because it is a product of disjoint swap operators.

For $n=1$, Eq.~\eqref{eq:observable-r2n} reduces to
\begin{equation}
    \hat O_2^{(r)}=\hat S_{A_1,A_2}\otimes\hat S_{B_1,B_2},
\end{equation}
so that
\begin{equation}
    r_2=\Tr\left[\hat\rho^{\otimes2}\left(\hat S_{A_1,A_2}\otimes\hat S_{B_1,B_2}\right)\right]=\Tr\left[\hat\rho^2\right],
\end{equation}
as expected.

\subsection{Interferometric implementation}

Although the initial expressions of $w_n$ and $r_{2n}$ involve the realignment map, this map does not need to be directly implemented physically. As a matter of fact, the corresponding multicopy observables can be measured using balanced beam splitters, homodyne detection, and photon-number-parity measurements, as detailed below.

\subsubsection{Measurement of $w_n$}

Let us consider $n$ copies of the bipartite state, namely,
\begin{equation}
    \hat\rho_{A_1B_1}\otimes\cdots\otimes\hat\rho_{A_nB_n}.
\end{equation}
Equation~\eqref{eq:wn-epr-projectors} shows that an EPR projection must be performed between Alice's subsystem of copy $t$ and Bob's subsystem of the preceding copy. The modes are therefore paired as
\begin{equation}
    (A_1,B_n),\quad(A_2,B_1),\quad\ldots,\quad(A_n,B_{n-1}).
\end{equation}
This pairing may be implemented by cyclically rerouting Bob's modes before the interferometers.

For each pair $(A_t,B_{t-1})$, with $B_0\coloneqq B_n$, we define the commuting EPR quadratures
\begin{equation}
    \hat x_{-,t}=\frac{\hat x_{B_{t-1}}-\hat x_{A_t}}{\sqrt{2}},\qquad\hat p_{+,t}=\frac{\hat p_{A_t}+\hat p_{B_{t-1}}}{\sqrt{2}}.
\end{equation}
Since they satisfy
\begin{equation}
    [\hat x_{-,t},\hat p_{+,t}]=0,
\end{equation}
their joint measurement can be implemented by a balanced beam splitter followed by homodyne detection on its two output modes.
The EPR projector can therefore be written~\cite{Hertz2021}
\begin{equation}
    \ket{\Omega}\!\bra{\Omega} = \pi \ket{p_+=0,x_-=0} \bra{p_+=0,x_-=0}.
\end{equation}

With the beam-splitter convention adopted in Eq.~\eqref{eq:beam_splitter_bis}, the relevant output quadratures obey
\begin{equation}
    \hat B_{XY}^\dagger \hat p_X \hat B_{XY}=\frac{\hat p_X+\hat p_Y}{\sqrt{2}}=\hat p_+,
\end{equation}
and
\begin{equation}
    \hat B_{XY}^\dagger \hat x_Y \hat B_{XY}=\frac{\hat x_Y-\hat x_X}{\sqrt{2}}=\hat x_-.
\end{equation}
where $\hat B_{XY}$ denotes the balanced beam splitter unitary between modes $X$ and $Y$. 
Thus, homodyne detection of $\hat p_X$ on the output mode of $X$ and $\hat x_Y$ on the output mode of $Y$ provides a simultaneous measurement of $\hat p_+$ and $\hat x_-$.

The EPR projector can consequently be written formally as
\begin{equation}
\label{eq:epr-homodyne}
\begin{aligned}
\ket{\Omega}\!\bra{\Omega}_{XY}
={}&\pi\hat B_{XY}^\dagger
\Big[
\left(\ket{p{=}0}\!\bra{p{=}0}\right)_X
\\
&\qquad\otimes
\left(\ket{x{=}0}\!\bra{x{=}0}\right)_Y
\Big]
\hat B_{XY}.
\end{aligned}
\end{equation}
Combining Eqs.~\eqref{eq:wn-epr-projectors} and~\eqref{eq:epr-homodyne}, we obtain
\begin{equation}
\label{eq:wn-homodyne-density}
    w_n=\pi^nP_{w_n}(\boldsymbol{0}),
\end{equation}
where $P_{w_n}(\boldsymbol{0})$ is the joint probability density for obtaining
\begin{equation}
    x_{-,t}=0,\qquad p_{+,t}=0,\qquad t=1,\ldots,n.
\end{equation}
For example, for low values of $n$, the interfering pairs on the balanced beam splitters are
\begin{equation}
    \begin{array}{c @{\hspace{0.65cm}} c}
    \hline
    n & \text{Alice-Bob interfering pairs}\\
    \hline
    1 & (A_1,B_1)\\
    2 & (A_1,B_2),\ (A_2,B_1)\\
    3 & (A_1,B_3),\ (A_2,B_1),\ (A_3,B_2).\\
    \hline
    \end{array}
\end{equation}
No genuine $n$-port interferometer is therefore required. The setup consists only of a cyclic rerouting followed by $n$ independent balanced beam splitters, all of which may operate in parallel, and ends with parallel homodyne measurements.
The corresponding setup for $w_3$ is illustrated in
the left panel of Fig.~\ref{fig:higher_order_implementation}.

\subsubsection{Measurement of $r_{2n}$}

The even moment $r_{2n}$ requires $2n$ copies of the state. According to Eq.~\eqref{eq:observable-r2n}, Alice's copies are interfered in the pairs
\begin{equation}
    (A_1,A_2),\quad(A_3,A_4),\quad\ldots,\quad(A_{2n-1},A_{2n}),
\end{equation}
while Bob's copies are interfered according to the staggered cyclic pairing
\begin{equation}
    (B_2,B_3),\quad(B_4,B_5),\quad\ldots,\quad(B_{2n},B_1).
\end{equation}

Each swap operator $\hat S_{XY}$ can be measured using a balanced beam splitter $\hat B_{XY}$ followed by a photon-number-parity measurement on the second output mode. With the convention of Eq.~\eqref{eq:beam_splitter_bis}, one has
\begin{equation}
\label{eq:swap-parity-general}
    \hat S_{XY}=\hat B_{XY}^{\dagger}(-1)^{\hat n_Y}\hat B_{XY},
\end{equation}
where $\hat n_Y$ denotes the photon-number operator of the second output mode.

Consequently,
\begin{equation}
\label{eq:r2n-parities}
    r_{2n}=\left\langle\prod_{t=1}^{n}(-1)^{\hat n_{A,t}^{\mathrm{out}}}\prod_{t=1}^{n}(-1)^{\hat n_{B,t}^{\mathrm{out}}}\right\rangle,
\end{equation}
where $\hat n_{A,t}^{\mathrm{out}}$ denotes the photon number in the second output mode of the beam splitter acting on
\begin{equation}
    (A_{2t-1},A_{2t}),
\end{equation}
and $\hat n_{B,t}^{\mathrm{out}}$ denotes the photon number in the second output mode of the beam splitter acting on
\begin{equation}
    (B_{2t},B_{2t+1}),\qquad B_{2n+1}\coloneqq B_1.
\end{equation}

Thus, $r_{2n}$ is obtained by multiplying the photon-number parity measured at the second output mode of all beam splitters and averaging the product over the experimental runs.

\begin{figure*}[t]
    \centering
    \includegraphics[width=0.95\textwidth]{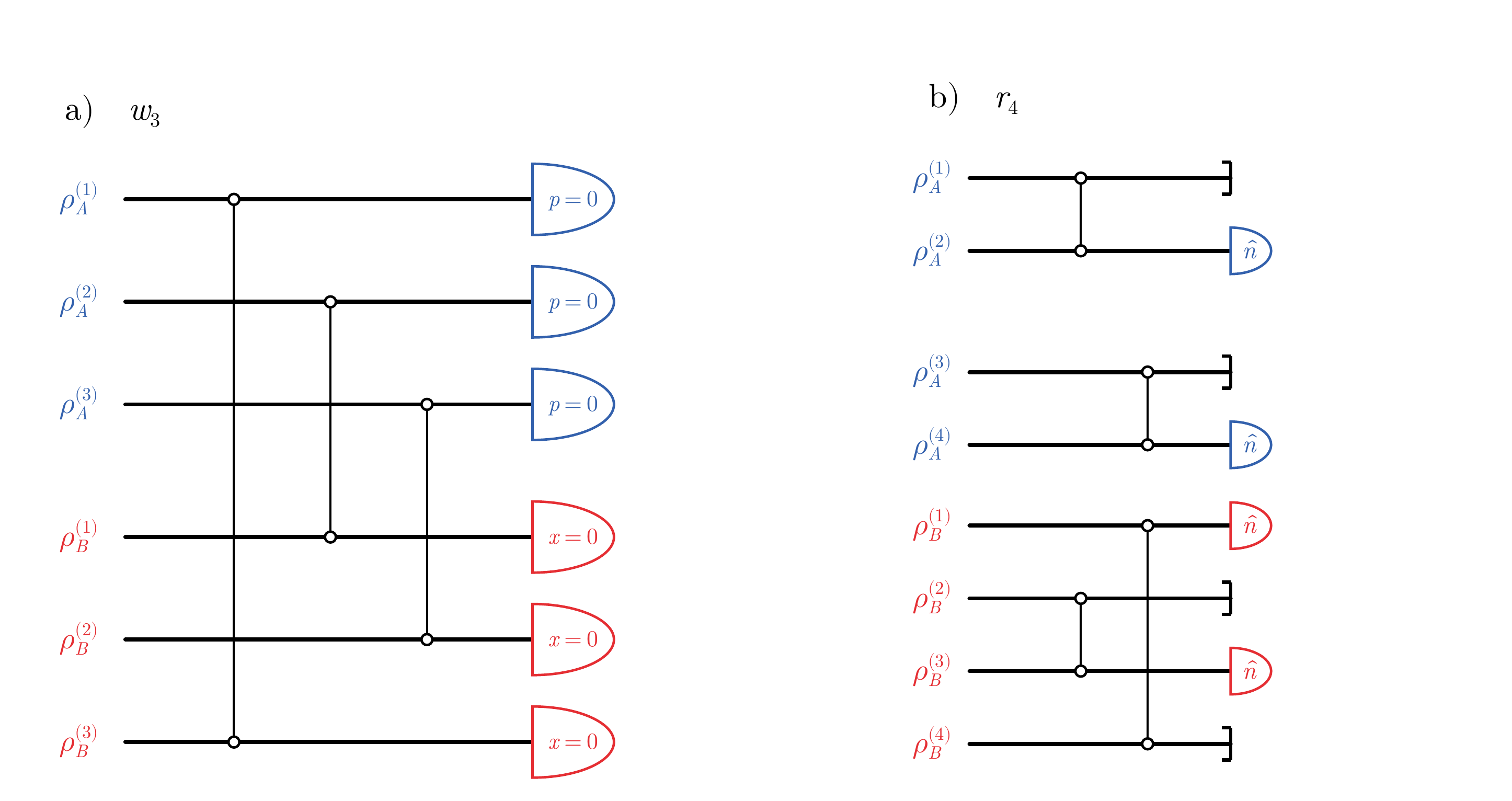}
    \caption{Interferometric measurement schemes for the higher-order moments $w_3$ and $r_4$. Vertical lines connecting two modes represent balanced beam splitters. \textbf{a)} Measurement of $w_3$ using three copies of the bipartite state. The beam splitters couple the pairs $(A_1,B_3)$, $(A_2,B_1)$, and $(A_3,B_2)$, followed by homodyne detection of the $p$ quadrature on Alice's output modes and the $x$ quadrature on Bob's output modes. \textbf{b)} Measurement of $r_4$ using four copies of the state. The beam splitters couple the pairs $(A_1,A_2)$ and $(A_3,A_4)$ on Alice's side, and $(B_2,B_3)$ and $(B_4,B_1)$ on Bob's side. Photon-number parity is measured at the second output mode of each beam splitter, while the other outputs are discarded. The product of the measured parities is averaged over the experimental runs.}
    \label{fig:higher_order_implementation}
\end{figure*}

For example, the purity $r_2$ requires two copies whereas fourth moment $r_4$ requires four copies, and the interfering pairs on the balanced beam splitters are 
\begin{equation}
    \begin{array}{c c @{\hspace{0.75cm}} c}
    \hline
    n & \text{Alice's pairs} & \text{Bob's pairs}\\
    \hline
    2 & (A_1,A_2) & (B_1,B_2)\\
    4 & (A_1,A_2),\ (A_3,A_4) & (B_2,B_3),\ (B_4,B_1). \\
    \hline
    \end{array}
\end{equation}
Here again, no genuine $n$-port interferometer is needed. The setup consists only of $n$ independent balanced beam splitters ($n/2$ on Alice's side and $n/2$ on Bob's side), all of which may operate in parallel, completed with $n$ parity measurements.
The corresponding setup for $r_4$ is illustrated in
the right panel of Fig.~\ref{fig:higher_order_implementation}.

\medskip

\section{Second-moment criterion and PPT}
\label{App:second-order-criterion+PPT}

 Here, we show that second-moment realignment criterion \eqref{eq:our_criterion} is weaker than the PPT criterion, hence it cannot detect bound entanglement. Specifically, we will prove that
\begin{equation}
    \hat\rho^{T_B} \geq 0 \implies r_2 + |w_2 - w_1^2| \leq 1.
    \label{eq:main}
\end{equation}
The proof goes as follows~\cite{MartinConversation}. 
Remember first that $r_2$ happens to coincide with the purity $\tr\left[\hat \rho^2\right]$, see Eq.~\eqref{eq:purity=r2}. This allows us to write
\begin{equation}
    \begin{aligned}
        1-r_2 &\stackrel{(a)}= \Tr[\hat{\rho}]^2-\Tr[\hat{\rho}^2]\\
        &\stackrel{(b)}=\Tr[\hat{\rho}^{T_B}]^2-\Tr[(\hat{\rho}^{T_B})^2]\\
        &\stackrel{(c)}=\Tr[\hat{\rho}^{T_B}\otimes \hat{\rho}^{T_B}]-\Tr[\hat{S}_{12}(\hat{\rho}^{T_B}\otimes \hat{\rho}^{T_B})]\\
        &\stackrel{(d)}=\Tr[(\hat\openone-\hat{S}_{12})(\hat{\rho}^{T_B}\otimes \hat{\rho}^{T_B})]\\
        &\stackrel{(e)}=\Tr[\hat Y],
    \end{aligned}
\end{equation}
where $(a)$ is by normalization of $\hat \rho$ and by definition of $r_2$, $(b)$ holds due to the partial transposition not affecting the two first moments of $\hat \rho$, $(c)$ is by definition of the trace of a tensor product and by Eq.~\eqref{eq:swapdef}, $(d)$ is by regrouping terms, and $(e)$ is the definition of the operator $\hat Y$. 

The antisymmetrization operator $(\hat\openone - \hat S_{12})/2$ is idempotent (it is a projector) and commutes with $\hat{\rho}^{T_B}\otimes \hat{\rho}^{T_B}$, so we can rewrite the operator $\hat Y$ as
\begin{equation}
    \hat Y = 2 \, \frac{\hat\openone - \hat{S}_{12}}{2}\, (\hat{\rho}^{T_B} \otimes \hat{\rho}^{T_B}) \, \frac{\hat\openone - \hat{S}_{12}}{2}.
\end{equation}
Thus, since we assume $\hat{\rho}^{T_B}$ to be positive semidefinite and since two-sided projections preserve positivity, we find $\hat Y\geq 0$.

Now, we express $w_1$ and $w_2$ by rewriting the realigned density matrix with the use of the partial transposition map, namely $R(\hat{\rho})=(\hat{\rho}^{T_B}\hat{S}_{AB})^{T_B}$~\cite{Hertz2021}. Following the same line of reasoning, we can write
\begin{equation}
    \begin{split}
        w_1^2-w_2 =&\Tr[(\hat{\rho}^{T_B}\hat{S}_{AB})^{T_B}]^2-\Tr[((\hat{\rho}^{T_B}\hat{S}_{AB})^{T_B})^2]\\ 
        =&\Tr[(\hat{\rho}^{T_B}\hat{S}_{AB})]^2-\Tr[(\hat{\rho}^{T_B}\hat{S}_{AB})^2]\\
        =&\Tr[(\hat\openone-\hat{S}_{12})\left((\hat{\rho}^{T_B}\hat{S}_{AB})\otimes(\hat{\rho}^{T_B}\hat{S}_{AB})\right)]\\
        =&\Tr[(\hat\openone-\hat{S}_{12})(\hat{\rho}^{T_B} \otimes \hat{\rho}^{T_B})(\hat{S}_{AB}\otimes \hat{S}_{AB})]\\
        =&\Tr[\hat Y \hat K],
    \end{split}
\end{equation}
where we define operator $K=\hat{S}_{AB}\otimes \hat{S}_{AB}$.

Since $K$ is a Hermitian unitary operator, we have
\begin{equation}
    -\hat\openone\leq \hat K \leq \hat\openone.
\end{equation}  
Using the positivity of $\hat Y$, it follows that
\begin{equation}
    -\hat Y\leq \hat Y^{\frac{1}{2}} \hat K \hat Y^{\frac{1}{2}} \leq \hat Y.
\end{equation}
Taking traces and using the invariance property of the trace under cyclic shifts yields
\begin{equation}
    |w_2 - w_1^2| = |\operatorname{Tr}[\hat{Y}\hat{K}]| \leq \operatorname{Tr}[\hat{Y}] = 1 - r_2,
    \label{eq:bound}
\end{equation}
which proves Eq.~\eqref{eq:main} in any local basis.

\bibliography{references.bib}

\end{document}